\documentclass[conference,compsoc]{IEEEtran}
\usepackage{float}
\usepackage{xcolor}
\usepackage{amssymb}
\usepackage{multicol}
\usepackage{multirow}
\usepackage{graphicx}

\usepackage[available,functional,reproduced]{ieeebadges}

\usepackage{colortbl}
\usepackage{makecell}
\usepackage{xspace}
\usepackage{listings}
\usepackage{amsmath}
\usepackage{booktabs}
\usepackage{fontawesome5}
\usepackage{enumitem}
\usepackage{booktabs}
\usepackage{threeparttable}
\usepackage{caption}
\usepackage{siunitx}
\usepackage{caption}
\usepackage{subcaption}

\usepackage{tikz}
\usepackage[most]{tcolorbox}

\definecolor{asrcolor0}{rgb}{0.75,0.93,0.81}
\definecolor{asrcolor1}{rgb}{0.81,0.94,0.79}
\definecolor{asrcolor2}{rgb}{0.88,0.96,0.78}
\definecolor{asrcolor3}{rgb}{0.94,0.97,0.76}
\definecolor{asrcolor4}{rgb}{1.00,0.99,0.75}
\definecolor{asrcolor5}{rgb}{0.99,0.92,0.73}
\definecolor{asrcolor6}{rgb}{0.99,0.84,0.71}
\definecolor{asrcolor7}{rgb}{0.98,0.77,0.69}

\definecolor{worsecolor}{rgb}{0.98,0.77,0.69}
\definecolor{comparablecolor}{rgb}{1.0, 0.95, 0.7}
\definecolor{bettercolor}{rgb}{0.75,0.93,0.81}
\newcommand{\worseperf}[1]{\cellcolor{worsecolor}#1}
\newcommand{\comparableperf}[1]{\cellcolor{comparablecolor}#1}
\newcommand{\betterperf}[1]{\cellcolor{bettercolor}#1}

\definecolor{lightred}{rgb}{1, 0.8, 0.8}         
\definecolor{lightyellow}{rgb}{1, 1, 0.8}        
\definecolor{lightgreen}{rgb}{0.8, 1, 0.8}     

\DeclareRobustCommand{\redcell}{\cellcolor{lightred}}
\DeclareRobustCommand{\yellowcell}{\cellcolor{lightyellow}}
\DeclareRobustCommand{\greencell}{\cellcolor{lightgreen}}

\DeclareRobustCommand{\yellowtext}[1]{\colorbox{lightyellow}{#1}}
\DeclareRobustCommand{\redtext}[1]{\colorbox{lightred}{#1}}
\DeclareRobustCommand{\greentext}[1]{\colorbox{lightgreen}{#1}}

\DeclareRobustCommand{\circletfill}{\tikz{\fill (0,0) circle (0.3em);}}
\DeclareRobustCommand{\circlet}{\tikz{\draw (0,0) circle (0.3em);}}

\DeclareRobustCommand{\cmark}{{\color{green}\checkmark}}%
\usepackage{pifont}
\usepackage{color}
\DeclareRobustCommand{\xmark}{{\color{red}\ding{55}}}
\DeclareRobustCommand{\human}{\raisebox{-0.1ex}{\footnotesize\faUser}}
\DeclareRobustCommand{\robot}{\raisebox{0.1ex}{\scriptsize\faRobot}}
\DeclareRobustCommand{\train}{\raisebox{-0.1ex}{\footnotesize\faCogs}}
\DeclareRobustCommand{\auto}{\raisebox{-0.1ex}{\footnotesize\faCog}}

\usepackage{tikz}
\newcommand*\circled[1]{\tikz[baseline=(char.base)]{
            \node[shape=circle,draw,inner sep=0.5pt] (char) {#1};}}
\ifCLASSOPTIONcompsoc
  \usepackage[nocompress]{cite}
\else
  \usepackage{cite}
\fi

\usepackage{adjustbox}

\ifCLASSINFOpdf
\else
\fi
\usepackage{url}
\usepackage[hidelinks]{hyperref}
\usepackage{bm}

\usepackage[dvipsnames]{xcolor}

\newtcolorbox{tkwtextbox}[1][]{%
    colback=black!5,
    colframe=black!5,
    notitle,
    sharp corners,
    borderline west={0pt}{0pt}{green!80!black},
    enhanced,
    breakable,
    left=0pt,
    right=0pt,
    top=0pt,
    bottom=0pt
    }

\newtcolorbox{opentextbox}[1][]{%
    colback=black!5,
    colframe=black!5,
    notitle,
    sharp corners,
    borderline west={0pt}{0pt}{red!80!black},
    enhanced,
    breakable,
    left=0pt,
    right=0pt,
    top=0pt,
    bottom=0pt
    }

\newcommand{\takeawayN}[1]{
    \begin{tkwtextbox}
        \textbf{Takeaway}: {#1}
    \end{tkwtextbox}
}

\newcommand{\openproblemN}[1]{
    \begin{opentextbox}
        \textbf{Open Problem}: {#1}
    \end{opentextbox}
}

\begin{document}

\title{SoK: Robustness in Large Language Models against Jailbreak Attacks}

\author{
\IEEEauthorblockN{Feiyue Xu\textsuperscript{1}, Hongsheng Hu\textsuperscript{1}, Chaoxiang He\textsuperscript{1}, Sheng Hang\textsuperscript{1}, Hanqing Hu\textsuperscript{1}, Xiuming Liu\textsuperscript{1}, Yubo Zhao\textsuperscript{1}, \\ Zhengyan Zhou\textsuperscript{1}, Bin Benjamin Zhu\textsuperscript{2}, Shi-Feng Sun\textsuperscript{1}, Dawu Gu\textsuperscript{1}, and Shuo Wang\textsuperscript{1}}
\IEEEauthorblockA{\textsuperscript{1}Shanghai Jiao Tong University, China}
\IEEEauthorblockA{\textsuperscript{2}Microsoft Corporation}
}

% make the title area
\maketitle
\thispagestyle{plain}
\pagestyle{plain}

\begin{abstract}
Large Language Models (LLMs) have achieved remarkable success but remain highly susceptible to jailbreak attacks, in which adversarial prompts coerce models into generating harmful, unethical, or policy-violating outputs. Such attacks pose real-world risks, eroding safety, trust, and regulatory compliance in high-stakes applications. Although a variety of attack and defense methods have been proposed, existing evaluation practices are inadequate, often relying on narrow metrics like attack success rate that fail to capture the multidimensional nature of LLM security.  
In this paper, we present a systematic taxonomy of jailbreak attacks and defenses and introduce $\mathtt{Security\; Cube}$, a unified, multi-dimensional framework for comprehensive evaluation of these techniques. We provide detailed comparison tables of existing attacks and defenses, highlighting key insights and open challenges across the literature. Leveraging $\mathtt{Security\; Cube}$, we conduct benchmark studies on 13 representative attacks and 5 defenses, establishing a clear view of the current landscape encompassing jailbreak attacks, defenses, automated judges, and LLM vulnerabilities. Based on these evaluations, we distill critical findings, identify unresolved problems, and outline promising research directions for enhancing LLM robustness against jailbreak attacks. Our analysis aims to pave the way towards more robust, interpretable, and trustworthy LLM systems. Our code is available at \href{https://github.com/XOTaichi/Security-Cube-Artifact}{\texttt{Code}}.
\end{abstract}

% For peer review papers, you can put extra information on the cover
% page as needed:
% \ifCLASSOPTIONpeerreview
% \begin{center} \bfseries EDICS Category: 3-BBND \end{center}
% \fi
%
% For peerreview papers, this IEEEtran command inserts a page break and
% creates the second title. It will be ignored for other modes.
\IEEEpeerreviewmaketitle

\section{Introduction}
\label{sec:introduction}
Large language models (LLMs) have rapidly become integral to a wide range of applications, including coding assistants~\cite{wang2021codet5, roziere2023code}, healthcare advice~\cite{cascella2023evaluating}, customer service~\cite{wulf2024exploring}, education~\cite{kasneci2023chatgpt}, and scientific research~\cite{zhang2025scientific}. Their strength stems from their ability to comprehend natural language, generate coherent and contextually appropriate responses, and adapt across diverse domains with remarkable flexibility. However, their growing adoption has sparked serious security concerns. In particular, malicious actors can jailbreak an LLM’s safety guardrails, coercing the model into producing prohibited or harmful outputs~\cite{jbchat, dan}. Early demonstrations revealed that even seemingly simple strategies, such as role-playing as an unconstrained agent or embedding hidden instructions, could circumvent safety policies in widely deployed models like GPT-3.5-Turbo~\cite{cipherchat}. 

Building on these initial findings, jailbreak attacks have since diversified and become increasingly sophisticated. Beyond early role-playing tricks, researchers have developed techniques such as gradient-based prompt optimization~\cite{gcg}, automated attack orchestration using auxiliary LLMs~\cite{pair, autodanturbo}, and adaptive strategies spanning multiple turns or modalities~\cite{actorattack,crescendo,Zhou2025TempestAM,niu2024jailbreaking}. As jailbreaks become more diverse and sophisticated, so too do their impacts: attacks now go beyond simple safety bypasses to produce contextually rich, targeted harmful outputs that enable more scalable and covert exploitation. In parallel, the community has proposed defenses such as adversarial training~\cite{ouyang2022training}, instruction alignment through fine-tuning or unlearning~\cite{saferl,safeunlearning}, and safety layers that filter outputs prior to delivery~\cite{aligner,autodefense}.

However, despite rapid progress in attack and defense techniques, evaluations of their effectiveness and robustness remain fragmented, often lacking standardized benchmarks and reproducible protocols. Existing frameworks~\cite{jbb, Shen2025PandaGuardSE} typically collapse performance into a single metric, attack success rate (ASR), while overlooking other critical dimensions such as stability across runs, cross-model transferability, and internal impact on model mechanisms. This one-dimensional perspective creates significant blind spots. For example, an attack may appear highly effective based on ASR yet fail to reproduce consistently, or two attacks with similar ASRs might exploit fundamentally different vulnerabilities, implying very different defense strategies. Such limitations risk misleading conclusions and impede the development of dependable safeguards. To address these gaps, a systematic study and a comprehensive evaluation framework are urgently needed to consolidate existing knowledge, illuminate trade-offs among current methods, and guide future research toward securing LLMs against jailbreak threats.

In this paper, we systematize the state of the art in jailbreak research for LLMs, providing a comprehensive landscape and taxonomy of attack strategies, defense mechanisms, and evaluation methodologies. Motivated by the limitations of existing benchmarks, such as their over-reliance on ASR and neglect of stability, transferability, and real-world considerations, we introduce $\mathtt{Security\; Cube}$, a multi-dimensional evaluation framework designed to capture a fuller spectrum of jailbreak characteristics. Leveraging this framework, we benchmark representative attacks and defenses, uncovering key insights into current challenges and highlighting opportunities for more robust and reliable LLM safety mechanisms. 

Our main contributions are summarized as follows:
\begin{itemize}[leftmargin=*]

\item \textbf{Comprehensive taxonomy of attacks and defenses}. 
We develop a systematic taxonomy that categorizes jailbreak attack and defense methods based on their dominant underlying mechanisms, ensuring clarity and minimal category overlap. Beyond the taxonomy, we derive key insights that provide a structured and comprehensive understanding of the jailbreak landscape.

\item \textbf{$\bm{\mathtt{\bm{Security\; Cube}}}$: a multi-dimensional evaluation framework}. 
We systematically review and compare existing evaluation benchmarks and identify critical gaps, such as their over-reliance on a single metric and narrow coverage across attack, defense, and judge types. To address these shortfalls, we propose $\mathtt{Security\;Cube}$, a holistic evaluation framework that integrates 7 attack metrics, 3 defense metrics, and 4 judge metrics, several introduced and evaluated for the first time. This design facilitates rigorous, comprehensive assessment of jailbreaks and defenses, establishing a stronger foundation for improving LLM security.
    
\item \textbf{Empirical benchmarking and key findings}. 
We apply $\mathtt{Security\;Cube}$ in a comprehensive study spanning 13 representative attacks, 5 defenses, and 4 automated judges across major categories. Our results uncover new vulnerabilities, reveal overlooked weaknesses, and demonstrate that multi-dimensional evaluations often contradict conclusions drawn from ASR alone.

\item \textbf{Research directions for LLM safety}. 
Building on our empirical findings, we outline promising directions to advance jailbreak research and LLM security, including exploring new attack and defense paradigms, instituting continuous red-teaming, developing community-shared benchmarks, and deepening understanding of jailbreak interpretation for safety assurance.
\end{itemize}

\section{Taxonomies of Attacks and Defenses}
\subsection{Problem Definition of Jailbreak}\label{sec:problemdef}

Let an LLM be a function $M(\cdot):\mathcal{P} \to \mathcal{R}$, where $\mathcal{P}$ is the space of prompts and $\mathcal R$ is the space of responses. An LLM jailbreak is an adversarial procedure that aims to induce a safety-aligned model \(M_{\mathrm{a}}(\cdot)\) to violate its safety policies.
Formally, given a harmful prompt \(P_{\mathrm{harm}}\) that would normally be refused by the aligned model \(M_{\mathrm{a}}(\cdot)\), the attacker constructs a jailbreak prompt \(P_{\mathrm{j}}\) through transformation $\mathcal T:\mathcal P\to\mathcal P$:
\[
  P_{\mathrm{j}} = \mathcal T(P_{\mathrm{harm}}).
\]
The model’s response to this jailbreak prompt is then:
\[
  R_{\mathrm{t}} = M_{\mathrm{a}}(P_{\mathrm{j}}).
\]
To assess whether the jailbreak is successful, we follow prior works~\cite{adaptive,boreiko2024realistic,jbb,pair} in employing a binary evaluation function \( \mathrm{Judge}(\cdot,\cdot): \mathcal{R} \times \mathcal{P} \to \{\mathsf{true}, \mathsf{false}\} \), which returns \( \mathsf{true} \) if the model’s output \( R_{\mathrm{t}} \) fulfills the malicious intent expressed in the original prompt \( P_{\mathrm{harm}} \).
Formally, a jailbreak is deemed successful if:
\[
  \mathrm{Judge}(R_{\mathrm{t}}, P_{\mathrm{harm}}) = \mathsf{true}.
\]
Note that \(P_{\mathrm{j}}\) is not used as input to the \(\mathrm{Judge}\) function, since the transformation \(\mathcal{T}\) may significantly alter the surface form of the prompt, making it unsuitable for intent matching.

\subsection{Threat Models}\label{sec:threat-model}
In this work, we systematically evaluate the robustness of LLMs against prompt-based jailbreak attacks by modeling a red-team adversary characterized as follows:

\noindent \textbf{Adversary goals.} The attacker aims to craft inputs that induce the model to violate safety policies and produce harmful content, while minimizing effort and maximizing attack generality across prompts, tasks, and models.

\noindent \textbf{Adversary capabilities.} We cover both black-box and white-box attack settings:
\begin{itemize}[leftmargin=*]
    \item \textbf{Black-box attackers} can access only the model’s input-output API, observe responses (including refusals), iteratively adapt prompts, and may use surrogate models.
    \item \textbf{White-box attackers} have full access to model internals (parameters, gradients, architecture) and can optimize adversarial prompts directly using the target model.
\end{itemize}

\noindent \textbf{Defender capabilities.} We also consider both black-box and white-box defense settings:

\begin{itemize}[leftmargin=*]
    \item \textbf{Black-box defenders} have access only to the model’s inputs and outputs. They can filter or modify the input before inference, and filter or adjust the output afterwards.
    \item \textbf{White-box defenders} have full access to the model internals. They can finetune the model or monitor its internal representations and parameters during inference.

\end{itemize}
\noindent \textbf{Constraints and assumptions.} We do not assume that the attacker has access to the training data, the ability to fine-tune the model, or the use of pre-filling techniques to elicit specific responses. This constraint ensures that the evaluation focuses on jailbreaks that arise from the model’s own behavior, rather than from external manipulations.

\subsection{Attack Overview}
To provide a structured view of the diverse landscape of red-teaming techniques against large language models, we propose a novel taxonomy of representative attack methods, summarized in Table~\ref{tab:redteam_sok}. In this taxonomy, each category is defined by a distinct underlying mechanism. For attack methods that could plausibly span multiple categories, we classify them according to their dominant mechanism.

\begin{table*}[!t]
\caption{Overview and comparison of existing jailbreak methods. \textbf{Code}: \cmark\xspace indicates open-source code is available; \xmark\xspace no open-source implementation. \textbf{Granularity}: \textit{T} means harmful prompt is created at the token level; \textit{P} at prompt level. \textbf{Access}: \circletfill\xspace black-box query access to the target model; \circletfill\xspace(S) black-box query access to the target model with a surrogate model; \circlet\xspace full white-box access. \textbf{Feedback}: \circletfill\xspace requires iterative feedback from the target model; \circlet\xspace no interaction needed. \textbf{Requirements}: \train\xspace means attackers require to fine-tune or train a red-teaming model; \robot\xspace means attackers employ an LLMs to generate harmful prompts; \circlet\xspace means attackers require only access to the target model. \textbf{Strategy}: \auto\xspace means the harmful prompt is automatically generated by specific attack algorithms or LLMs; \human\xspace means the harmful prompt is mainly generated by human designed rules, e.g., template, with LLMs assisting in tasks like rewriting or role-playing. \textbf{Diversity} measures the complexity of variants in the harmful prompt. \greentext{Low} indicates the variants of the harmful prompt for a single goal is fewer than 10; \yellowtext{Medium} indicates moderate number of variants between 10-100. \redtext{High} indicates more than 100 variants, i.e, a broad and significant variation in wording, syntax, and attack logic.}

\centering
\small
\resizebox{0.8\textwidth}{!}{
\begin{tabular}{lcccccccccc} 
\toprule
\textbf{Type} & \textbf{Attack}           & \textbf{Reference}        & \textbf{Year} & \textbf{Code}               & \textbf{Granularity}      & \textbf{Access}              & \textbf{Feedback}       &  \textbf{Requirements}             & \textbf{Strategy}   & \textbf{Diversity}               \\ 
\midrule
\multirow{9}{*}{\centering Logprob} 
& GCG & \cite{gcg} & 2023 & \cmark & T & \circlet\xspace & \circletfill\xspace  & \train  & \auto  & \redcell Low  \\ 
& AutoDAN & \cite{autodan} & 2023 & \cmark & T+P & \circlet\xspace & \circletfill\xspace  & \train  & \auto  & \yellowcell Medium \\ 
& AmpleGCG & \cite{amplegcg} & 2024 & \cmark & T & \circlet\xspace & \circletfill\xspace  & \train  & \auto  & \redcell Low  \\ 
& ColdAttack & \cite{cold} & 2024 & \cmark & T+P & \circlet\xspace & \circletfill\xspace  & \train  & \auto  & \yellowcell Medium \\ 
& DSN & \cite{dsn} & 2024 & \cmark & T & \circlet\xspace & \circletfill\xspace  & \train  & \auto  & \redcell Low  \\ 
& MAC & \cite{mac} & 2024 & \cmark & T & \circlet\xspace & \circletfill\xspace  & \train & \auto  & \redcell Low  \\ 
& PAL & \cite{pal} & 2024 & \cmark & T & \circletfill\xspace (S)  & \circletfill\xspace  & \train \xspace \robot & \auto  & \redcell Low  \\ 
& LLM-Adaptive & \cite{adaptive} & 2024 & \cmark & T+P & \circletfill\xspace (S)  & \circletfill\xspace  & \train \xspace \robot & \auto  & \yellowcell Medium \\ 
& DualBreach & \cite{dualbreach} & 2025 & \xmark & T & \circletfill\xspace (S)  & \circletfill\xspace  & \train \xspace \robot & \auto  & \redcell Low  \\ \midrule

\multirow{2}{*}{\centering Shuffle} & BON & \cite{bon} & 2024 & \cmark & T & \circletfill\xspace  & \circletfill\xspace  & \robot(J) & \auto  & \yellowcell Medium \\ 
 & Flip & \cite{flip} & 2025 & \cmark & T & \circletfill\xspace  & \circlet\xspace & \circlet\xspace & \human & \redcell Low  \\ \midrule
 
\multirow{6}{*}{\centering LLM} 
& PAIR & \cite{pair} & 2023 & \cmark & P & \circletfill\xspace  & \circletfill\xspace  & \robot & \auto  & \greencell High \\ 
& TAP & \cite{tap} & 2023 & \cmark & P & \circletfill\xspace  & \circletfill\xspace  & \robot & \auto  & \greencell High \\ 
& AutoDAN-Turbo & \cite{autodanturbo} & 2024 & \cmark & P & \circletfill\xspace  & \circletfill\xspace  & \train \xspace \robot & \auto \xspace \human & \greencell High \\
& JailPo & \cite{Li2024JailPOAN} & 2024 & \xmark & P & \circletfill\xspace & \circletfill\xspace  & \robot & \auto & \greencell High \\ 
& ArrAttack & \cite{arrattack} & 2025 & \xmark & P & \circletfill\xspace  & \circlet\xspace & \train \xspace\robot & \auto  & \greencell High \\ 
& Auto-RT & \cite{Liu2025AutoRTAJ} & 2025 & \cmark & P & \circletfill\xspace  & \circlet\xspace & \train \xspace\robot & \auto\xspace \human   & \greencell High \\ \midrule

\multirow{7}{*}{\centering Strategy}  & ReNeLLM & \cite{renellm} & 2023 & \cmark & P & \circletfill\xspace  & \circletfill\xspace  & \robot & \human & \yellowcell Medium \\ 
& PAP & \cite{pap} & 2024 & \cmark & P & \circletfill\xspace  & \circlet\xspace & \robot & \human & \yellowcell Medium \\ 
 & DAN & \cite{dan} & 2024 & \cmark & P & \circletfill\xspace  &\circlet\xspace  & \circlet\xspace & \human & \redcell Low  \\ 
 & CodeAttacker & \cite{codeattack} & 2024 & \cmark & P & \circletfill\xspace  &\circlet\xspace  & \circlet\xspace & \human & \redcell Low  \\ 
 & QueryAttack & \cite{queryattack} & 2025 & \cmark & P & \circletfill\xspace  &\circlet\xspace  & \circlet\xspace & \human & \yellowcell Medium \\ 
 & DrAttack & \cite{drattack} & 2024 & \cmark & P & \circletfill\xspace  & \circletfill\xspace  & \robot & \human & \yellowcell Medium \\ 
 & DIE & \cite{die} & 2025 & \xmark & P & \circletfill\xspace  & \circletfill\xspace  & \robot & \human & \yellowcell Medium \\ 
 & ICRT & \cite{icrt} & 2025 & \cmark & P & \circletfill\xspace  & \circletfill\xspace  & \robot & \human & \yellowcell Medium \\ \midrule
 
\multirow{5}{*}{\centering Multi-round} 
& DeepInception & \cite{deepinception} & 2023 & \cmark & P & \circletfill\xspace  & \circletfill\xspace  & \robot & \human & \yellowcell Medium \\ 
& Crescendo & \cite{crescendo} & 2024 & \xmark & P & \circletfill\xspace  & \circletfill\xspace  & \circlet\xspace   & \human & \greencell High \\ 
 
 & MRJ & \cite{mrj} & 2024 & \xmark & P & \circletfill\xspace  & \circletfill\xspace  & \robot & \auto  & \greencell High \\ 
 & ActorBreaker & \cite{actorattack} & 2024 & \cmark & P & \circletfill\xspace  & \circletfill\xspace  & \robot & \auto  & \greencell High \\
 & Tempest & \cite{Zhou2025TempestAM} & 2025 & \cmark &
 P & \circletfill\xspace  & \circletfill\xspace  & \robot & \auto  & \greencell High \\\midrule
 
\multirow{3}{*}{\centering Flaw} & CipherChat & \cite{cipherchat} & 2023 & \cmark & P & \circletfill\xspace  & \circlet\xspace &  \circlet\xspace& \human & \redcell Low  \\ 
 & Multijail & \cite{multilingual} & 2023 & \cmark & P & \circletfill\xspace  & \circlet\xspace & \circlet\xspace & \human & \redcell Low  \\ 
 & ArtPrompt & \cite{art} & 2024 & \cmark & P & \circletfill\xspace  & \circlet\xspace & \circlet\xspace & \human & \redcell Low  \\ \midrule
\centering Template & GPTFuzzer & \cite{gptfuzzer} & 2023 & \cmark & P & \circletfill\xspace  & \circletfill\xspace  & \robot & \human & \yellowcell Medium \\ 
\bottomrule
\end{tabular}
}
\label{tab:redteam_sok}

\end{table*}

\noindent\textbf{Attack taxonomy}. Specifically, we categorize existing attack methods into seven types:
\begin{itemize}[leftmargin=*]
    \item \textbf{Logprobe-based attacks} refer to attacks that exploit exposed diagnostic signals, e.g., log outputs, auxiliary classifier scores, or (in white-box cases) gradient/activation information, to iteratively choose inputs that raise the chance the model produces disallowed content.
    \item \textbf{Shuffle-based attacks} try to evade safety by making harmless textual edits, e.g., reordering tokens, inserting separators or extra whitespace, or other small surface-level perturbations, so the prompt remains semantically intact for humans but changes how the model parses or responds.
    \item \textbf{LLM-based attacks} refer to attacks that use other LLMs as the primary means to automatically create, paraphrase, or iteratively refine prompts with the goal of increasing the chance a target model will produce disallowed content.
    \item \textbf{Multi-round-based (conversational) attacks} use multiple dialogue turns to gradually steer the model into producing disallowed content by building context, lowering guard signals, or exploiting state carried across turns.
    \item \textbf{Flaw-based attacks} exploit specific, model-dependent weaknesses, e.g., quirks in multilingual handling, failure modes from pretraining data gaps, or systematic miscalibrations, to craft inputs that reliably cause unsafe outputs.
    \item \textbf{Strategy-based attacks} refer to attacks that rely primarily on human-designed manipulative strategies, e.g., social, procedural, or rhetorical devices, to coax the model into producing disallowed content. For example, attackers use persuasive or procedural structure (framing, step requests, or persona appeals) to make an unsafe outcome seem like a legitimate, harmless task.
    \item \textbf{Template-based attacks} start from a set of complete, pre-defined prompts and then generate many variants by applying controlled mutations (e.g., phrase permutation). The attacker searches these mutated versions to find prompts that bypass safety.

\end{itemize}

For each attack method we summarize key features (granularity, access, feedback, requirements, strategy, diversity) in Table~\ref{tab:redteam_sok}, producing a taxonomy of red‑teaming (e.g., manual vs. automated) that clarifies core mechanisms and enables systematic comparison of strengths and limitations.

\noindent \textbf{Key observations.} From Table~\ref{tab:redteam_sok}, we gain the following key observations of the current jailbreak attack landscape:

\noindent \textbf{\circled{1} Logprob, Strategy, and LLM-based attacks dominate the LLM jailbreak landscape.} 
When categorized by method, most jailbreaks since 2023 cluster into three classes: logprob-based, strategy-based, and LLM-based attacks, with other classes (shuffle, multi-round, flaw, and template-based methods) appearing less frequently. Specifically, Logprob-based attacks thrive because exposed diagnostics give highly informative feedback, making search efficient. Strategy-based attacks succeed since they exploit universal tendencies of LLMs to be helpful and follow instructions. LLM-based attacks dominate by automating generation and refinement, enabling scale and creativity with little manual effort. Overall, these three approaches dominate because they combine effectiveness, generalizability, and ease of execution, making them more practical than  alternatives.

\noindent \textbf{\circled{2} Jailbreak threats in practical black-box attack settings.}
From the access and granularity perspective, the majority of attacks primarily involve prompt-level manipulations that operate in black-box settings, relying solely on input–output access to the target model. These attacks are impactful because they require no insider access or gradients: an adversary only needs the public API or chat interface to probe, iterate, and exploit behavioral weaknesses, which makes them easy to scale and automate. In addition, such attacks operate at the prompt level in a more computation-friendly manner than token-level approaches, i.e., they need fewer, coarser queries and less compute per candidate, lowering cost and latency of the attack, and thus posing a broad, low-barrier threat in practice.

\noindent\textbf{\circled{3} Red-teaming challenges using existing jailbreaks.} Despite their prevalence, most jailbreak attacks still rely heavily on human-guided strategies, which often produce low-diversity prompts that recycle fixed patterns and repetitive structures, making them easier to detect and less robust. At the same time, more advanced approaches, such as those logprob-based attacks involving optimization or external LLM orchestration, face significant trade-offs in computation and API cost, limiting scalability and reproducibility. Together, these factors show that existing jailbreak techniques are not yet fully automated, struggle to generate diverse adversarial prompts, and remain expensive to scale, challenges that make comprehensive and systematic red-teaming of LLMs difficult.

\subsection{Defense Overview}
\begin{table*}[t]
\caption{Overview and comparison of defense methods against jailbreak attacks.
\textbf{Code}: \cmark\xspace indicates open-source code is available; \xmark\xspace no open-source implementation. \textbf{Access}: \circletfill\xspace black-box access to the target model; \circlet\xspace white-box access. \textbf{Flexibility}: \cmark\xspace means the defense is model-agnostic and transferable across LLMs; \xmark\xspace model-specific defense. \textbf{Extra LLM}: \robot\xspace an LLM is required; \circlet\xspace LLMs are not required; \textbf{Fine-tuning requirement}: \train\xspace requires fine-tuning procedures, either aligning the target model itself or fine-tuning an alternative model for defense; \circlet\xspace fine-tuning is not needed; \textbf{Cost}: \yellowtext{Training} indicates the cost of defense is at the fine-tuning time of the target model; \redtext{Iterative} indicates the cost is at the inference time of the target model, with multiple rounds of inference to secure against a potential harmful prompt; \greentext{Single-pass} indicates cost is at the inference time of the target model with only one round for defense.
}

\centering
% \small
% \setlength{\tabcolsep}{0.6em}
% \renewcommand{\arraystretch}{1.1}
\resizebox{0.8\textwidth}{!}{
\begin{tabular}{lccccccccc} 
\toprule
\textbf{Type} & \textbf{Defense}           & \textbf{Reference}        & \textbf{Year} & \textbf{Code}                    & \textbf{Access}              & \textbf{Flexibility}      & \textbf{Extra LLM}         & \textbf{Fine-tuning}   & \textbf{Cost}              \\ 
\midrule 
\multirow{4}{*}{\centering Pre-filter} & Erase-and-check & \cite{erase} & 2023 & \cmark & \circletfill\xspace  & \cmark & \circlet\xspace & \circlet\xspace & \redcell Iterative \\ 
 & LlamaGuard & \cite{llamaguard} & 2023 & \cmark & \circletfill\xspace  & \cmark & \robot & \train & \greencell Single-pass \\ 
 & Perplexity Filter & \cite{perplexity} & 2023 & \xmark & \circletfill\xspace  & \cmark & \robot & \train & \greencell Single-pass \\ 
  & Hidden State Guard & \cite{hidden} & 2024 & \cmark & \circletfill\xspace  & \cmark & \robot & \train & \greencell Single-pass \\ \midrule

System prompt & SelfRemider & \cite{reminder} & 2023 & \cmark & \circletfill\xspace  & \cmark & \circlet\xspace  & \circlet\xspace & \greencell Single-pass \\ \midrule

\multirow{7}{*}{\centering Fine-tune} & Constitutional AI & \cite{constitutional} & 2022 & \xmark & 
\circlet\xspace & \xmark & \circlet\xspace & \train & \yellowcell Training \\ 
& Safe-RLHF & \cite{saferl} & 2023 & \cmark & \circlet\xspace & \xmark & \circlet\xspace  & \train & \yellowcell Training \\ 
 & Deep Alignment & \cite{tokendeep} & 2024 & \xmark & \circlet\xspace & \xmark & \circlet\xspace  & \train & \yellowcell Training \\ 
 & Safety-Tuned LLaMAs & \cite{tune} & 2024 & \cmark & \circlet\xspace & \xmark & \circlet\xspace & \train & \yellowcell Training \\ 
 & Safe unlearning & \cite{safeunlearning} & 2024 & \cmark & \circlet\xspace & \xmark & \circlet\xspace & \train & \yellowcell Training \\ 
 & CircuitBreaker & \cite{circuitbreaker} & 2024 & \cmark & \circlet\xspace & \xmark & \circlet\xspace & \train & \yellowcell Training \\ 
 & Scot & \cite{scot} & 2025 & \cmark & \circlet\xspace &  \xmark & \circlet\xspace & \train & \yellowcell Training \\ 
 \midrule

\multirow{4}{*}{\centering Intra-process} & PAT & \cite{pat} & 2024 & \cmark & \circlet\xspace &  \xmark & \circlet\xspace & \train & \greencell Single-pass \\ 
 & ABD & \cite{abd} & 2024 & \xmark & \circlet\xspace &  \xmark & \circlet\xspace & \train  & \greencell Single-pass \\ 
 & SafeDecoding & \cite{safedecoding} & 2024 & \cmark & \circlet\xspace & \cmark & \robot & \train & \greencell Single-pass \\   
 & JBShield & \cite{jbshield} & 2025 & \cmark & \circlet\xspace &  \xmark & \circlet\xspace & \train & \greencell Single-pass \\\midrule
\multirow{4}{*}{\centering Post-filter}
 & Smoothllm & \cite{smoothllm} & 2023 & \cmark & \circletfill\xspace  &  \cmark & \circlet\xspace  & \circlet\xspace & \redcell Iterative \\ 
 & Aligner & \cite{aligner} & 2024 & \cmark & \circlet\xspace & \cmark& \robot & \train & \greencell Single-pass \\  
 & Self-evaluation & \cite{selfeval} & 2024 & \cmark & \circletfill\xspace  & \cmark & \circlet\xspace  & \circlet\xspace & \greencell Single-pass \\ 
 & Autodefense & \cite{autodefense} & 2024 & \cmark & \circletfill\xspace  & \cmark & \robot & \circlet\xspace  & \redcell Iterative \\ 

\bottomrule
\end{tabular}
}
\label{tab:defense_sok}

\end{table*}

To systematically examine defense strategies against jailbreaks in large language models (LLMs), we propose a taxonomy organized by the stage of defense deployment, as summarized in Table~\ref{tab:defense_sok}. This framework provides a structured overview that facilitates comparison of defenses across the model life cycle, while also highlighting the practical trade-offs associated with each approach.

\noindent \textbf{Defense taxonomy}. We categorize defense methods into five types based on their deployment stage in the model pipeline:
\begin{itemize}[leftmargin=*]
    \item \textbf{Pre-filter defenses} work by screening user inputs before they reach the target model. The idea is to detect and block potentially malicious or jailbreak-inducing prompts at the entry point, which is usually implemented with auxiliary classifiers or with manually crafted rules.
    \item \textbf{System prompt defenses} strengthen safety by embedding explicit guardrails into the model’s system prompt. This involves inserting safety instructions or constraints that steer the model’s behavior during inference.
    \item \textbf{Fine-tuning-based defenses} improve robustness by retraining or further aligning the model itself on curated data. Instead of filtering inputs or relying on prompts, the model’s parameters are updated so that it inherently resists malicious instructions and produces safer outputs.
    \item \textbf{Intra-process defenses} intervene during inference by modifying or monitoring the model’s decoding process. Techniques may include constraining token sampling, biasing logits toward safe outputs, or interrupting decoding when unsafe continuations are detected.
    \item \textbf{Post-filter defenses}  operate after the model has generated its response, filtering, revising, or blocking unsafe outputs before they reach the user. This approach can use classifiers, rule-based filters, or rewriting mechanisms to detect and suppress problematic content.
\end{itemize}

\noindent \textbf{Key observations.} From Table~\ref{tab:defense_sok}, we summarize the following key observations of the current defense methods:

\noindent \textbf{\circled{1} White-box fine-tuning–based defenses dominate.}
Fine-tuning is one of the most prevalent strategies because it offers a fundamental and durable form of robustness: once a model has been fine-tuned on carefully curated safety data, it can resist a wide range of attacks without requiring ongoing external interventions. This makes fine-tuning attractive as a long-term solution, since it internalizes safety into the model parameters and incurs minimal inference-time cost. However, this white-box dependency introduces significant scalability challenges. That is, each model must undergo its own fine-tuning or alignment process, which demands substantial compute, expertise, and high-quality data. As a result, while fine-tuning is powerful and deeply integrated, its lack of portability and heavy upfront overhead limit its flexibility in multi-model or black-box settings.

\noindent \textbf{\circled{2} High computational cost of defenses.}
Many existing defense methods introduce substantial computational overhead, either during training or at inference time. Approaches based on fine-tuning demand significant upfront resources for retraining and alignment, raising development costs and limiting scalability. Meanwhile, other defenses that avoid fine-tuning often shift the burden to inference: requiring multiple forward passes or auxiliary model calls to safeguard against a single potentially harmful query. This can multiply inference costs, degrade latency, and complicate deployment in real-world systems where efficiency is critical. Ultimately, this highlights a central tension: while strong defenses are desirable, the cost-benefit trade-off between robust safety guarantees and efficient service remains unresolved.

\noindent\textbf{\circled{3} Defensive efforts lag far behind attacks.} As shown in Table~\ref{tab:redteam_sok}, the number and diversity of defensive studies are considerably smaller than that of attack methods. This imbalance arises in part because attacks are inherently easier to devise: adversaries need only find a single exploitative pathway, whereas defenders must anticipate and guard against a broad and evolving range of threats. The implications of this gap are significant. If attack research continues to outpace defenses, LLMs may remain persistently vulnerable, with new jailbreak techniques regularly undermining existing safeguards. This asymmetry highlights the need for sustained investment in proactive and generalizable defense strategies, as well as better benchmarks to track progress and close the gap between offensive and defensive innovation.

\section{Evaluation Framework}

\subsection{$\mathtt{Security\; Cube}$ }\label{sec:cubeframe}
\begin{figure*}[t]
  \centering
  \includegraphics[width=1\linewidth]{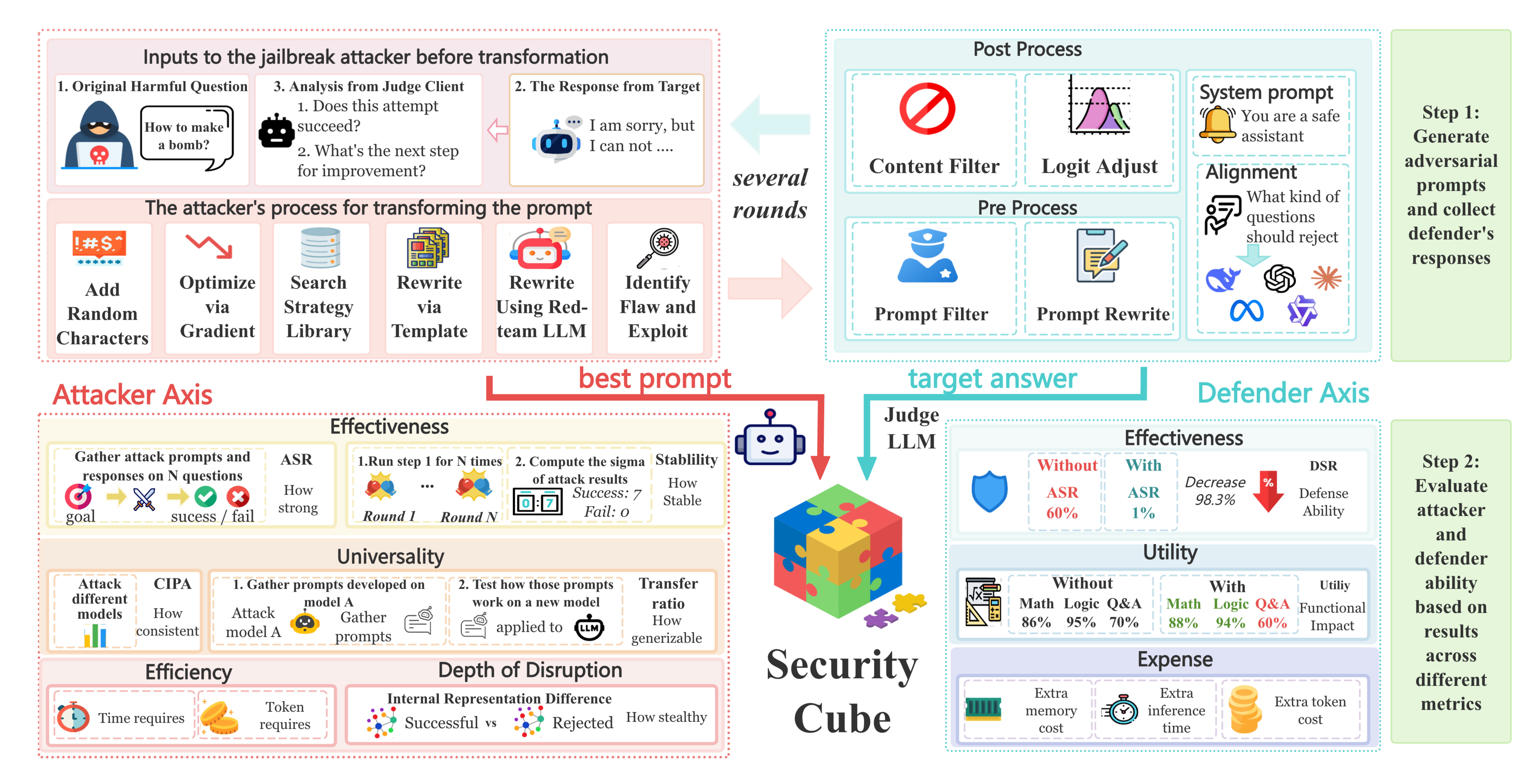}
  \caption{Overview of the $\mathtt{Security\;Cube}$ pipeline. Given a jailbreak goal, the attacker generates an initial adversarial prompt using a specific attack method (e.g., shuffling, LLM-based generation, or template rewriting). The target model, protected by a defense mechanism such as system prompts, pre-/post-guardrails, or other safety layers, produces a response. The attacker iteratively refines the prompt based on defender feedback (black-box or white-box), applying early stopping and incorporating suggestions. The final effective prompt–response pair is evaluated by a Judge model to assess attack success. Throughout the process, $\mathtt{Security\;Cube}$ logs key metrics of the attack, defense, and judge components.}
  \label{fig:pipeline}
\end{figure*}

In this section, we introduce $\mathtt{Security\; Cube}$, a novel multi-dimensional framework (Figure~\ref{fig:pipeline}) for evaluating LLM robustness and safety along three axes: attacker, defender, and judge. Unlike prior benchmarks that examine isolated aspects, $\mathtt{Security\; Cube}$ unifies 7 attack, 3 defense, and 3 judge metrics, enabling a systematic mapping of the jailbreak landscape. Crucially, it introduces three new metrics, attack stability, concentration index per attack (CIPA), and depth of disruption, providing the first formal measures of adversarial consistency, concentration, and cascading impact. These innovations fill critical gaps in existing evaluations, offering a more comprehensive and structured assessment than prior frameworks. Section~\ref{sec:related_work} and Table~\ref{tab:benchmark_sok} further contrasts $\mathtt{Security\; Cube}$ with existing benchmarks.

\noindent\textbf{Attacker axis} evaluates a jailbreak attack across five dimensions, success, stability, transferability, disruption depth, and overhead, offering a quantitative, multidimensional basis for comparison, detection, and mitigation.

\begin{itemize}[leftmargin=*]

\item \textbf{Attack success rate (ASR)}, denoted by $\alpha$, the fraction of adversarial prompts that bypass a model’s safety mechanisms. Given a harmful prompt $P$, a jailbreak prompt $P_j$ is constructed and the model produces a response $R$. A binary evaluation function $\mathrm{Judge}(R, P_j)$ assigns $S=1$ if the attempt is successful (i.e., the model violates its safety policy) and $S=0$ otherwise. Over $N$ total attempts, ASR is defined as 
\[
\alpha = \frac{1}{N} \sum_{k=1}^{N} S_k,
\]
where larger $\alpha$ indicates more effective jailbreaks.

\item \textbf{Attack stability}, denoted by $\beta$, measures the consistency of an attack in bypassing a model’s safety mechanisms under similar conditions (e.g., fixed temperature, varying random seeds). For a given harmful prompt $P$, a specific attack generates $N$ repeated trials, yielding an attack success rate $\alpha$. Over $K$ distinct harmful prompts, this produces success rates $\alpha_1, \ldots, \alpha_K$. The attack stability is then defined as the standard deviation of these success rates:
\[\beta = \mathrm{std}(\alpha_1, \ldots, \alpha_K),\]
where lower values of $\beta$ indicate higher stability, i.e., the jailbreak attack is more reliably effective across prompts.

\item \textbf{Attack transferability.} This metric family assesses the transferability of jailbreak attacks, with three concrete metrics define as follows:

\begin{itemize}[leftmargin=1em] % Adjusted leftmargin for nested list
\item \textbf{Cross-model transferability}, denoted by $\eta$, quantifies how well an attack crafted for one model transfers to another model. Formally, consider $N$ adversarial prompts $P_i$ generated by a particular attack with respect to a source model. When these prompts are evaluated on a target model, let $S_i \in \{0,1\}$ indicate whether the $i$-th prompt successfully bypasses the target model’s safety mechanisms. The cross-model transferability is then defined as
\[
\eta = \frac{1}{N}\sum_{i=1}^{N} S_i,
\]
where higher values of $\eta$ indicate that the attack generalizes more effectively across models.

\item \textbf{Transfer ratio}, denoted by $\gamma$, measures the relative effectiveness of an attack on a target model compared to its effectiveness on the source model. Formally, let $\alpha$ denote the attack success rate on the source model and $\eta$ the attack success rate on the target model (i.e., cross-model transferability). The transfer ratio is defined as
\[
\gamma = \frac{\eta}{\alpha},
\]
where values of $\gamma$ close to 1 indicate that the attack transfers nearly as effectively as it works on the source model, values below 1 indicate weaker transfer to the target, and values above 1 (though uncommon in practice) indicate that the attack is even more effective on the target than on the source.

\item \textbf{Concentration Index per Attack (CIPA).} Inspired by the Herfindahl–Hirschman Index~\cite{hhi}, we define CIPA to quantify how concentrated or generalizable the success of an attack is across $N$ different victim models. Let $\alpha_1, \ldots, \alpha_N$ denote the attack success rates of the same attack on $N$ models. Then CIPA is given by
\[
\text{CIPA} = \sum_{i=1}^N (\frac{\alpha_i}{\sum_{j=1}^{N} \alpha_j})^2,
\]
where higher values indicate that the attack’s success is concentrated on only a few models, while lower values indicate that the attack generalizes more broadly across different models.

\end{itemize}

Together, the three transferability metrics provide a multi-dimensional view of jailbreak robustness. Cross-model transferability and transfer ratio capture the portability of attacks, highlighting systemic risks when vulnerabilities easily transfer across models. CIPA reflects attack diversity, distinguishing between broad, generalizable jailbreaks and narrow, model-specific exploits. Particularly concerning are attacks with low CIPA and high transferability, as they are both widely effective and transferable, whereas high CIPA with low transferability suggests more isolated and easier-to-mitigate weaknesses.

\item \textbf{Depth of disruption}, denoted by $\mu$, measures the representational divergence between successful and unsuccessful jailbreak prompts within a model. Let $\mathcal{D}=\{P_1,\ldots,P_n\}$ be a dataset of prompts generated by a specific attack, partitioned into successful jailbreaks $\mathcal{D}_s$ and failures $\mathcal{D}_f$. Denote by $h(P)$ the hidden representation of prompt $P$ at a given model layer. The depth of disruption at that layer is defined as
\[
\mu = \mathrm{dist}(\frac{1}{|\mathcal{D}_s|} \sum_{P \in \mathcal{D}_s} h(P), \; \frac{1}{|\mathcal{D}_f|} \sum_{P \in \mathcal{D}_f} h(P)),
\]
where $\mathrm{dist}(\cdot,\cdot)$ denotes a distance metric such as cosine distance. Higher values of $\mu$ indicate stronger representational divergence, reflecting more severe disruption of the model’s internal understanding caused by the attack.

\item \textbf{Attack overhead} quantifies the computational cost of generating jailbreak prompts, measured in terms of both token consumption and elapsed time across $N$ interaction rounds. The total token cost is defined as
\[
T_{\text{token}} = \sum_{i=1}^{N} \left( T^{(i)}_{\text{prompt}} + T^{(i)}_{\text{output}} \right),
\]
where $T^{(i)}_{\text{prompt}}$ and $T^{(i)}_{\text{output}}$ denote the number of input and output tokens, respectively, for round $i$. This includes tokens consumed by both the target model and any auxiliary red-teaming models employed during the attack. The total time cost is defined as
\[
T_{\text{time}} = \sum_{i=1}^{N} T^{(i)},
\]
where $T^{(i)}$ is the wall-clock time (in seconds) taken for the $i$-th interaction round. Higher values of attack overhead indicate greater resource demands, which may limit the practicality of a jailbreak in real-world settings.

\end{itemize}

\noindent\textbf{Defender axis} evaluates a jailbreak defense across three dimensions, defense effectiveness, overhead, and utility preservation, offering a balanced view of how well the defense mitigates attacks, its computational cost, and its impact on normal model functionality.

\begin{itemize}[leftmargin=*]

\item \noindent \textbf{Defense success rate (DSR)}, denoted by $\nu$, measures how effectively a defense reduces the success of jailbreak attacks, serving as the counterpart to ASR. It is defined as the relative reduction in attack success rate after applying a defense. Let $\alpha$ denote the original attack success rate on the undefended model, and let $\alpha_d$ denote the attack success rate after applying the defense. The defense success rate is defined as
\[
\nu = \frac{\alpha - \alpha_d}{\alpha},
\]
where $\nu = 0$ indicates that the defense has no effect, and $\nu = 1$ indicates that the defense completely prevents all successful attacks. Higher values of DSR therefore correspond to more effective defenses.

\item \textbf{Utility preservation ability} measures the extent to which a defense maintains the model’s performance on standard tasks, thereby quantifying potential utility degradation. Let $\mathcal{D}$ be a dataset of clean (non-adversarial) tasks. The relative utility change is defined as
\[
\Delta U = U^{\text{defense}}(\mathcal{D}) - U^{\text{undefended}}(\mathcal{D}),
\]
where $U^{\text{defense}}$ and $U^{\text{undefended}}$ denote task performance (e.g., accuracy) with and without the defense, respectively. Values of $\Delta U$ close to $0$ indicate strong utility preservation, while negative values indicate degradation in the model’s normal utility due to the defense.

\item \textbf{Defense overhead} quantifies the additional computational cost introduced by a defense mechanism, measured in terms of token usage, time, and memory. The token overhead is defined as
\[
\Delta T_{\text{token}} = T_{\text{token}}^{\text{defense}} - T_{\text{token}}^{\text{undefended}},
\]
where $T_{\text{token}}^{\text{defense}}$ and $T_{\text{token}}^{\text{undefended}}$ denote the total token consumption with and without the defense, respectively. Similarly, the time and memory overhead are defined as
\[
\Delta T_{\text{time}} = T_{\text{time}}^{\text{defense}} - T_{\text{time}}^{\text{undefended}},\]
\[\Delta M = M^{\text{defense}} - M^{\text{undefended}}.
\]
A defense is considered practical when $\Delta T_{\text{token}}$, $\Delta T_{\text{time}}$, and $\Delta M$ remain small, ensuring that security improvements do not come at excessive computational or economic cost.

\end{itemize}

\begin{table*}[t]
\caption{Comparison of our work with existing benchmarks, frameworks, and surveys on jailbreaks against LLMs. Each work is systematically compared across four key dimensions: attacks, defenses, judges, and evaluated target models. From the attacker perspective, we consider ASR, stability, transferability, disruption, and overhead. From the defender perspective, we assess DSR, utility preservation, and overhead. The judge perspective captures disagreement with humans, agreement consistency, and evaluation cost. Finally, the target model dimension covers variations in model size and version. \circletfill\; and \circlet\; represent cover and not cover, respectively.}
\centering
\resizebox{0.85\linewidth}{!}{
\begin{tabular}{cccccccccccccccc}
\toprule
\multicolumn{1}{c}{}                               & \multicolumn{1}{c}{}                                 & \multicolumn{1}{c}{}                                  & \multicolumn{5}{c}{\cellcolor[HTML]{FFFFFF}\textbf{Attacker Perspective}} & \multicolumn{3}{c}{\cellcolor[HTML]{FFFFFF}\textbf{Defender Perspective}}                                         & \multicolumn{3}{c}{\cellcolor[HTML]{FFFFFF}\textbf{Judge Perspective}}                                                     & \multicolumn{2}{c}{\cellcolor[HTML]{FFFFFF}\textbf{Target Model}}                  \\ \cmidrule(lr){4-8} \cmidrule(lr){9-11} \cmidrule(lr){12-14} \cmidrule(lr){15-16}

\multicolumn{1}{c}{\multirow{-2}{*}{\textbf{Paper}}} & \multicolumn{1}{c}{\multirow{-2}{*}{\textbf{Year}}} & \multicolumn{1}{c}{\multirow{-2}{*}{\textbf{Type}}} & \multicolumn{1}{c}{\textbf{ASR}} & \multicolumn{1}{c}{\textbf{Stability}} & \multicolumn{1}{c}{\textbf{Transferability}} & \multicolumn{1}{c}{\textbf{Disruption}} & \multicolumn{1}{c}{\textbf{Overhead}} & \multicolumn{1}{c}{\textbf{DSR}} & \multicolumn{1}{c}{\textbf{Utility-preservation}} & \multicolumn{1}{c}{\textbf{Overhead}} & \multicolumn{1}{c}{\textbf{Disagreement}} & \multicolumn{1}{c}{\textbf{Consistency}} & \multicolumn{1}{c}{\textbf{Cost}} & \multicolumn{1}{c}{\textbf{Size}} & \multicolumn{1}{c}{\textbf{Version}} \\
\midrule
\cite{esmradi2023comprehensive} & 2023 & Survey & \circletfill\xspace & \circlet\xspace & \circlet\xspace & \circlet\xspace & \circlet\xspace & \circlet\xspace & \circlet\xspace & \circlet\xspace & \circlet\xspace & \circlet\xspace & \circlet\xspace & \circlet\xspace & \circlet\xspace \\
\cite{gcg} & 2023 & Benchmark & \circletfill\xspace & \circlet\xspace & \circletfill\xspace & \circlet\xspace & \circlet\xspace & \circlet\xspace & \circlet\xspace & \circlet\xspace & \circlet\xspace & \circlet\xspace & \circlet\xspace & \circlet\xspace & \circletfill\xspace \\
\cite{xu2024comprehensive} & 2024 & Survey & \circletfill\xspace & \circlet\xspace & \circlet\xspace & \circlet\xspace & \circletfill\xspace & \circletfill\xspace & \circletfill\xspace & \circlet\xspace & \circletfill\xspace & \circletfill\xspace & \circlet\xspace & \circlet\xspace & \circlet\xspace \\
\cite{xubag} & 2024 & Benchmark & \circletfill\xspace & \circlet\xspace & \circlet\xspace & \circlet\xspace & \circlet\xspace & \circlet\xspace & \circlet\xspace & \circlet\xspace & \circlet\xspace & \circlet\xspace & \circlet\xspace & \circletfill\xspace & \circletfill\xspace \\
\cite{boreiko2024realistic} & 2024 & Threat model & \circletfill\xspace & \circlet\xspace & \circlet\xspace & \circlet\xspace & \circlet\xspace & \circlet\xspace & \circlet\xspace & \circlet\xspace & \circletfill\xspace & \circletfill\xspace & \circlet\xspace & \circletfill\xspace & \circletfill\xspace \\
\cite{harmbench} & 2024 & Benchmark & \circletfill\xspace & \circlet\xspace & \circlet\xspace & \circlet\xspace & \circlet\xspace & \circletfill\xspace & \circlet\xspace & \circlet\xspace & \circletfill\xspace & \circlet\xspace & \circlet\xspace & \circletfill\xspace & \circletfill\xspace \\
\cite{AA} & 2024 & Survey & \circletfill\xspace & \circlet\xspace & \circlet\xspace & \circlet\xspace & \circlet\xspace & \circletfill\xspace & \circlet\xspace & \circlet\xspace & \circletfill\xspace & \circlet\xspace & \circlet\xspace & \circlet\xspace & \circletfill\xspace \\
\cite{yang2024benchmarking} & 2024 & Benchmark & \circletfill\xspace & \circlet\xspace & \circlet\xspace & \circlet\xspace & \circlet\xspace & \circletfill\xspace & \circletfill\xspace & \circlet\xspace & \circlet\xspace & \circlet\xspace & \circlet\xspace & \circlet\xspace & \circlet\xspace \\
\cite{mou2024sg} & 2024 & Benchmark & \circletfill\xspace & \circlet\xspace & \circlet\xspace & \circlet\xspace & \circlet\xspace & \circletfill\xspace & \circlet\xspace & \circlet\xspace & \circlet\xspace & \circlet\xspace & \circlet\xspace & \circletfill\xspace & \circletfill\xspace \\
\cite{h4rm3l} & 2024 & Benchmark & \circletfill\xspace & \circlet\xspace & \circletfill\xspace & \circletfill\xspace & \circlet\xspace & \circlet\xspace & \circlet\xspace & \circlet\xspace & \circletfill\xspace & \circlet\xspace & \circlet\xspace & \circletfill\xspace & \circletfill\xspace \\
\cite{zhou2024easyjailbreak} & 2024 & Framework & \circletfill\xspace & \circlet\xspace & \circlet\xspace & \circlet\xspace & \circletfill\xspace & \circlet\xspace & \circlet\xspace & \circletfill\xspace & \circletfill\xspace & \circletfill\xspace & \circletfill\xspace & \circletfill\xspace & \circletfill\xspace \\
\cite{munoz2024pyrit} & 2024 & Framework & \circletfill\xspace & \circlet\xspace & \circlet\xspace & \circlet\xspace & \circlet\xspace & \circlet\xspace & \circlet\xspace & \circlet\xspace & \circlet\xspace & \circlet\xspace & \circlet\xspace & \circlet\xspace & \circlet\xspace \\
\cite{derczynski2024garak} & 2024 & Framework & \circletfill\xspace & \circlet\xspace & \circlet\xspace & \circlet\xspace & \circlet\xspace & \circlet\xspace & \circlet\xspace & \circlet\xspace & \circlet\xspace & \circlet\xspace & \circlet\xspace & \circlet\xspace & \circlet\xspace \\
\cite{Surur2024JailbreakAO} & 2024 & Survey & \circletfill\xspace & \circlet\xspace & \circlet\xspace & \circlet\xspace & \circletfill\xspace & \circletfill\xspace & \circlet\xspace & \circlet\xspace & \circlet\xspace & \circlet\xspace & \circlet\xspace & \circlet\xspace & \circlet\xspace \\
\cite{decodingtrust} & 2024 & Benchmark & \circletfill\xspace & \circlet\xspace & \circlet\xspace & \circlet\xspace & \circlet\xspace & \circlet\xspace & \circlet\xspace & \circlet\xspace & \circlet\xspace & \circlet\xspace & \circlet\xspace & \circlet\xspace & \circletfill\xspace \\
\cite{trustllm} & 2024 & Benchmark & \circletfill\xspace & \circlet\xspace & \circlet\xspace & \circlet\xspace & \circlet\xspace & \circlet\xspace & \circlet\xspace & \circlet\xspace & \circlet\xspace & \circlet\xspace & \circlet\xspace & \circletfill\xspace & \circlet\xspace \\
\cite{jailbreakeval} & 2024 & Framework & \circletfill\xspace & \circlet\xspace & \circlet\xspace & \circlet\xspace & \circlet\xspace & \circlet\xspace & \circlet\xspace & \circlet\xspace & \circletfill\xspace & \circlet\xspace & \circlet\xspace & \circlet\xspace & \circlet\xspace \\
\cite{jbb} & 2024 & Benchmark & \circletfill\xspace & \circlet\xspace & \circlet\xspace & \circlet\xspace & \circletfill\xspace & \circletfill\xspace & \circlet\xspace & \circlet\xspace & \circletfill\xspace & \circlet\xspace & \circlet\xspace & \circlet\xspace & \circletfill\xspace \\
\cite{Rao2023TrickingLI} & 2024 & Survey & \circletfill\xspace & \circlet\xspace & \circlet\xspace & \circletfill\xspace & \circlet\xspace & \circlet\xspace & \circlet\xspace & \circlet\xspace & \circletfill\xspace & \circletfill\xspace & \circlet\xspace & \circlet\xspace & \circlet\xspace \\
\cite{Yu2024DontLT} & 2024 & Framework & \circletfill\xspace & \circlet\xspace & \circlet\xspace & \circlet\xspace & \circlet\xspace & \circlet\xspace & \circlet\xspace & \circlet\xspace & \circlet\xspace & \circlet\xspace & \circlet\xspace & \circlet\xspace & \circlet\xspace \\
\cite{peng2024jailbreaking} & 2025 & Survey & \circletfill\xspace & \circlet\xspace & \circletfill\xspace & \circlet\xspace & \circletfill\xspace & \circletfill\xspace & \circletfill\xspace & \circletfill\xspace & \circlet\xspace & \circlet\xspace & \circlet\xspace & \circlet\xspace & \circlet\xspace \\
\cite{thu} & 2025 & Framework & \circletfill\xspace & \circlet\xspace & \circlet\xspace & \circlet\xspace & \circlet\xspace & \circletfill\xspace & \circlet\xspace & \circlet\xspace & \circletfill\xspace & \circlet\xspace & \circlet\xspace & \circlet\xspace & \circlet\xspace \\
\cite{Chu2024JailbreakRadarCA} & 2025 & Benchmark & \circletfill\xspace & \circlet\xspace & \circlet\xspace & \circlet\xspace & \circletfill\xspace & \circletfill\xspace & \circlet\xspace & \circlet\xspace & \circletfill\xspace & \circletfill\xspace & \circlet\xspace & \circlet\xspace & \circlet\xspace \\
\cite{Shen2025PandaGuardSE} & 2025 & Benchmark & \circletfill\xspace & \circlet\xspace & \circlet\xspace & \circlet\xspace & \circletfill\xspace & \circletfill\xspace & \circlet\xspace & \circletfill\xspace & \circletfill\xspace & \circletfill\xspace & \circlet\xspace & \circletfill\xspace & \circletfill\xspace \\
\cite{Ma2025SafetyAS} & 2025 & Survey & \circletfill\xspace & \circlet\xspace & \circlet\xspace & \circlet\xspace & \circlet\xspace & \circletfill\xspace & \circlet\xspace & \circlet\xspace & \circlet\xspace & \circlet\xspace & \circlet\xspace & \circlet\xspace & \circlet\xspace \\
\cite{zizzo2025adversarial} & 2025 & Benchmark & \circletfill\xspace & \circlet\xspace & \circlet\xspace & \circlet\xspace & \circlet\xspace & \circletfill\xspace & \circletfill\xspace & \circletfill\xspace & \circlet\xspace & \circlet\xspace & \circlet\xspace & \circlet\xspace & \circlet\xspace \\
\cite{mai2025you} & 2025 & Benchmark & \circletfill\xspace & \circlet\xspace & \circlet\xspace & \circlet\xspace & \circlet\xspace & \circletfill\xspace & \circletfill\xspace & \circletfill\xspace & \circlet\xspace & \circlet\xspace & \circlet\xspace & \circlet\xspace & \circletfill\xspace \\
\cite{safedialbench} & 2025 & Benchmark & \circletfill\xspace & \circlet\xspace & \circlet\xspace & \circlet\xspace & \circlet\xspace & \circlet\xspace & \circlet\xspace & \circlet\xspace & \circletfill\xspace & \circlet\xspace & \circlet\xspace & \circletfill\xspace & \circletfill\xspace \\
\midrule
\textbf{Ours} & 2025 & SoK & \circletfill\xspace & \circletfill\xspace & \circletfill\xspace & \circletfill\xspace & \circletfill\xspace & \circletfill\xspace & \circletfill\xspace & \circletfill\xspace & \circletfill\xspace & \circletfill\xspace & \circletfill\xspace & \circletfill\xspace & \circletfill\xspace \\
\bottomrule
\end{tabular}
}
\label{tab:benchmark_sok}
\end{table*}

\noindent \textbf{Evaluator axis} systematically evaluates automated judges by assessing their annotation quality against human experts and measuring their scalability in practice. Annotation quality is captured through agreement-based metrics (e.g., disagreement with humans, inter-annotator agreement), while evaluation cost quantifies the computational resources required for large-scale use.

\begin{itemize}[leftmargin=*]

\item \textbf{Disagreement with humans} measures the reliability of an automated judge~\cite{xu2024comprehensive, jailbreakeval, jbb} by comparing its outputs against human annotations. Let $\alpha_h$ and $\alpha_j$ denote the attack success rates (ASRs) computed using human annotations and the automated judge, respectively. The disagreement is quantified as the absolute ASR bias:
\[
\Delta \text{ASR} = \left| \alpha_h - \alpha_j \right|.
\]
Lower values of $\Delta \text{ASR}$ indicate closer alignment with human judgment (and thus higher reliability), while higher values reflect greater disagreement, suggesting the automated judge is less trustworthy.

\item \textbf{Inter-annotator agreement ($\kappa$)} uses Fleiss' kappa~\cite{fleiss1971measuring} to evaluate the consistency of annotations across human annotators and automated judges in labeling jailbreak attempts. Let $N$ be the total number of jailbreak attempts, $n$ the number of annotators, and $k=2$ the number of categories (success or fail). Denote by $n_{ij}$ the number of annotators assigning the $i$-th attempt to category $j$. The agreement for attempt $i$ is
\[
A_i = \frac{1}{n(n-1)} \sum_{j=1}^{k} n_{ij}(n_{ij}-1),
\]
and the mean observed agreement is
\[
\bar{A} = \frac{1}{N} \sum_{i=1}^{N} A_i.
\]
Let $a_j = \frac{1}{Nn} \sum_{i=1}^{N} n_{ij}$ denote the marginal probability of category $j$, and define the expected agreement as
\[
\bar{A}_e = \sum_{j=1}^{k} a_j^2.
\]
Fleiss’ kappa is then given by
\[
\kappa = \frac{\bar{A} - \bar{A}_e}{1 - \bar{A}_e},
\]
which accounts for chance agreement. Higher values of $\kappa$ indicate stronger consistency among annotators, with $\kappa=1$ representing perfect agreement, $\kappa=0$ chance-level agreement, and negative values indicating systematic disagreement among annotators.

\item \textbf{Evaluation cost} quantifies the computational expense of using an automated judge, measured as the average token consumption per evaluation. Formally, for $N$ evaluations, the cost is defined as
\[
\tilde{T}_{\text{token}} = \frac{1}{N} \sum_{i=1}^{N} \left( T_{i}^{\text{prompt}} + T_{i}^{\text{output}} \right),
\]
where $T_{i}^{\text{prompt}}$ and $T_{i}^{\text{output}}$ denote the input and output tokens consumed by the employed LLM(s) in the $i$-th evaluation. Lower values of $\tilde{T}_{\text{token}}$ indicate more cost-efficient automated judging.

\end{itemize}

\begin{table*}[t]
\centering
\caption{Summary of LLMs used for jailbreak evaluation in $\mathtt{Security\; Cube}$.}
\label{tab:llm_features}
\resizebox{0.9\linewidth}{!}{
\begin{tabular}{lccccccc}
\toprule
\textbf{Model} & \textbf{Year} & \textbf{Size} & \textbf{Open / Closed Source} & \textbf{Access} & \textbf{Instr.~Tuned} & \textbf{Alignment Method} \\
\midrule
GPT-3.5-Turbo\cite{gpt3}         & 2023 & Not disclosed & Closed & API only & Yes & Not disclosed \\
Qwen-2.5-7B-Instruct\cite{qwen25max}  & 2024       & 7B           & Open (Apache-2.0) & Both & Yes & Online RL \\
LLaMA-3-8B-Instruct\cite{llama3}   & 2024       & 8B           & Open (llama3 license) & Both & Yes & Guardrails, Data Cleaning(pretrain), SFT, DPO\cite{llama3} \\
Mistral-7B-Instruct-v0.2\cite{mistral7b}   & 2023       & 7B           & Open (Apache-2.0) & Both & Yes & System Prompt\cite{mistral7b} \\
o1-mini~\cite{o1}     & 2024      & Not disclosed     & Closed & API only  & Yes & Guardrails, Data Filtering(pretrain), Deliberative Alignment \cite{o1}\\
DeepSeek-v3~\cite{deepseek} & 2024 & 631B & Open & Both & Yes & SafeRL\cite{deepseekr1} \\
Qwen-2.5-Max~\cite{qwen25max} & 2025 & Not disclosed & Closed & API only & Yes & SFT, RL\cite{qwen25max-report} \\
Gemini-2.0-Flash\cite{gemini2} & 2025 & Not disclosed & Closed & API only & Yes & Guardrails, Data filtering(pretrain), SFT, RLHF \cite{gemini2}
\\
Claude-3.7-Sonnet\cite{anthropic}     & 2025 & Not disclosed & Closed & API only & Yes & Guardrails, RLHF\cite{anthropic}\\
Qwen3-235B-A22B~\cite{qwen3} & 2025 & 235B & Open (Apache-2.0) & Both & Yes &  Not disclosed\\

\bottomrule
\end{tabular}
}
\end{table*}

\subsection{Comparison to Existing Works}\label{sec:related_work}

We present a systematic comparison of our work with 28 evaluation benchmarks, frameworks, and surveys on LLM jailbreaks in Table~\ref{tab:benchmark_sok}. The table highlights key characteristics across four dimensions: attacks, defenses, judges, and target models. From this comparison, we identify several major limitations of existing efforts as follows.

\noindent \textbf{\circled{1} Limited metric coverage.} A major limitation in existing works is their over-reliance on a narrow set of metrics. Specifically, 20 out of 28 surveyed studies (over 70\%) evaluate jailbreak attacks in LLMs solely through attack success rate. This singular focus limits understanding of how attacks generalize across diverse scenarios, and fails to capture the adaptive nature of real-world adversaries. Moreover, drawing conclusions based only on success rates risks misleading interpretations. For example, an attack that achieves a high success rate in one setting may not consistently succeed across different runs or contexts. Another issue is the widespread use of automated judges, such as AlpacaEval~\cite{alpaca_eval} and MT-Bench~\cite{mtbench101}, to evaluate attacks and defenses. These frameworks typically restrict evaluation to a few coarse metrics (e.g., ASR or DSR). However, such metrics can be gamed. For instance, a trivial defense that always produces safe-looking but uninformative outputs can still achieve high scores and be deemed ``robust''~\cite{zheng2024cheating}.

\noindent \textbf{\circled{2} Limited attacks, defenses, and target model coverage.} Another major shortcoming of existing benchmarks lies in their limited scope across attacks, defenses, and target models. Many benchmarks cover only a narrow range of attack types and defenses, and evaluate on a small set of victim models. For example, some focus exclusively on prompt-based attacks~\cite{safedialbench}, while others fail to test defenses across multiple model families or versions~\cite{jbb}. This limited coverage is particularly problematic given the rapid pace of LLM development, where models are continuously released with stronger alignment and safety mechanisms. As a result, an attack that was once considered effective may no longer succeed on newer versions, even within the same model family. Without broader evaluation, benchmarks risk presenting outdated or incomplete insights.

Overall, our proposed $\mathtt{Security\; Cube}$ addresses these gaps by broadening coverage across metrics, attacks, defenses, and target models, thereby capturing richer aspects of jailbreak dynamics. In doing so, it enables more systematic and reliable assessment of robustness in real-world, fast-evolving LLM ecosystems.

\section{$\mathtt{Security\; Cube}$ Evaluation}

\begin{table*}[t]
    \centering
    \caption{Attack success rate (ASR, in \%) of different jailbreak methods across model families. Cooler shades (green) denote low ASR (0–37.5\%, higher safety), yellow indicates moderate ASR (37.5–62.5\%), and warmer shades (red) denote high ASR (62.5–100\%, lower safety).}
     \label{tab:asr_models_updated}
    \resizebox{0.75\linewidth}{!}{%
        \begin{threeparttable}
            \begin{tabular}{lcccccccccc} 
                \toprule
                % & \multicolumn{4}{c}{\textbf{Old Models}} & \multicolumn{6}{c}{\textbf{New Models}} \\
                % \cmidrule(lr){2-5} \cmidrule(lr){6-11}
                \textbf{Attack Method} &
                \multicolumn{1}{c}{\textbf{GPT-3.5}} & \multicolumn{1}{c}{\textbf{Qwen}} & \multicolumn{1}{c}{\textbf{Llama}} & \multicolumn{1}{c}{\textbf{Mistral}} &
                \multicolumn{1}{c}{\textbf{Qwen2.5-Max}} & \multicolumn{1}{c}{\textbf{DeepSeek}} & \multicolumn{1}{c}{\textbf{o1-mini}} & \multicolumn{1}{c}{\textbf{Claude-3-7}} & 
                \multicolumn{1}{c}{\textbf{Gemini-2.0}} & \multicolumn{1}{c}{\textbf{Qwen3-235b}} \\
                \midrule
                LLM-Adaptive & \cellcolor{asrcolor7} 95.0 & \cellcolor{asrcolor7} 96.0 & \cellcolor{asrcolor7} 97.0 & \cellcolor{asrcolor7} 91.0 & \cellcolor{asrcolor0} 0.0 & \cellcolor{asrcolor7} 99.0 & \cellcolor{asrcolor0} 0.0 & \cellcolor{asrcolor0} 5.0 &  \cellcolor{asrcolor0} 0.0 & \cellcolor{asrcolor7} 95.0 \\
                ActorBreaker  & \cellcolor{asrcolor5} 70.0 & \cellcolor{asrcolor6} 79.5 & \cellcolor{asrcolor5} 68.0 & \cellcolor{asrcolor6} 76.0 & \cellcolor{asrcolor5} 73.0 & \cellcolor{asrcolor5} 69.5 & \cellcolor{asrcolor4} 53.0 & \cellcolor{asrcolor2} 35.5 & \cellcolor{asrcolor1} 25.0 & \cellcolor{asrcolor5} 64.5 \\
                BON           & \cellcolor{asrcolor3} 42.5 & \cellcolor{asrcolor1} 20.0 & \cellcolor{asrcolor0} 11.0 & \cellcolor{asrcolor1} 23.0 & \cellcolor{asrcolor2} 26.5 & \cellcolor{asrcolor3} 38.0 & \cellcolor{asrcolor0} 3.5  & \cellcolor{asrcolor0} 4.5  & \cellcolor{asrcolor0} 4.0 & \cellcolor{asrcolor0} 10.0 \\
                Flip          & \cellcolor{asrcolor4} 62.0 & \cellcolor{asrcolor3} 49.0 & \cellcolor{asrcolor1} 19.5 & \cellcolor{asrcolor2} 26.0 & \cellcolor{asrcolor4} 51.0 & \cellcolor{asrcolor6} 80.5 & \cellcolor{asrcolor0} 0.0 & \cellcolor{asrcolor2} 26.5  & \cellcolor{asrcolor2} 27.5 & \cellcolor{asrcolor5} 73.5 \\
                PAIR          & \cellcolor{asrcolor5} 67.5 & \cellcolor{asrcolor4} 51.0 & \cellcolor{asrcolor2} 37.0 & \cellcolor{asrcolor6} 79.5 & \cellcolor{asrcolor3} 47.5 & \cellcolor{asrcolor4} 54.0 & \cellcolor{asrcolor1} 16.5 & \cellcolor{asrcolor1} 14.0 & \cellcolor{asrcolor4} 54.5 & \cellcolor{asrcolor6} 80.0 \\
                AutoDAN-Turbo & \cellcolor{asrcolor3} 40.0 & \cellcolor{asrcolor4} 52.0 & \cellcolor{asrcolor3} 40.0 & \cellcolor{asrcolor5} 62.0 & \cellcolor{asrcolor2} 31.5 & \cellcolor{asrcolor4} 55.0 & \cellcolor{asrcolor1} 19.5 & \cellcolor{asrcolor1} 18.5 & \cellcolor{asrcolor0} 11.5 & \cellcolor{asrcolor6} 72.0 \\
                ReNeLLM       & \cellcolor{asrcolor6} 83.0 & \cellcolor{asrcolor6} 82.0 & \cellcolor{asrcolor5} 66.0 & \cellcolor{asrcolor7} 91.0 & \cellcolor{asrcolor0} 0.0 & \cellcolor{asrcolor6} 86.5 & \cellcolor{asrcolor5} 68.5 & \cellcolor{asrcolor6} 79.5 & \cellcolor{asrcolor2} 27.0 & \cellcolor{asrcolor6} 81.0 \\
                DrAttack      & \cellcolor{asrcolor2} 26.5 & \cellcolor{asrcolor2} 37.0 & \cellcolor{asrcolor1} 15.5 & \cellcolor{asrcolor1} 25.0 & \cellcolor{asrcolor4} 49.0 & \cellcolor{asrcolor3} 49.5 & \cellcolor{asrcolor0} 2.5  & \cellcolor{asrcolor0} 2.5  & \cellcolor{asrcolor0} 3.0 & \cellcolor{asrcolor4} 46.0 \\
                Multijail     & \cellcolor{asrcolor5} 68.5 & \cellcolor{asrcolor2} 35.0 & \cellcolor{asrcolor3} 47.0 & \cellcolor{asrcolor3} 48.0 & \cellcolor{asrcolor1} 18.0 & \cellcolor{asrcolor1} 21.0 & \cellcolor{asrcolor0} 2.0  & \cellcolor{asrcolor1} 17.0 & \cellcolor{asrcolor1} 13.0 & \cellcolor{asrcolor2} 28.0 \\
                CipherChat    & \cellcolor{asrcolor0} 9.5  & \cellcolor{asrcolor1} 14.0 & \cellcolor{asrcolor1} 20.5 & \cellcolor{asrcolor5} 68.0 & \cellcolor{asrcolor1} 13.0 & \cellcolor{asrcolor3} 40.0 & \cellcolor{asrcolor0} 0.5  & \cellcolor{asrcolor3} 42.5 & \cellcolor{asrcolor1} 15.5 & \cellcolor{asrcolor3} 50.0 \\
                CodeAttacker    & \cellcolor{asrcolor3} 49.5 & \cellcolor{asrcolor4} 58.0 & \cellcolor{asrcolor4} 57.5 & \cellcolor{asrcolor6} 73.0 & \cellcolor{asrcolor5} 69.5 & \cellcolor{asrcolor4} 56.0 & \cellcolor{asrcolor2} 33.0 & \cellcolor{asrcolor3} 40.0 & \cellcolor{asrcolor2} 28.0 & \cellcolor{asrcolor6} 80.0 \\
                PAP           & \cellcolor{asrcolor5} 62.5 & \cellcolor{asrcolor6} 75.0 & \cellcolor{asrcolor3} 39.0 & \cellcolor{asrcolor4} 60.0 & \cellcolor{asrcolor4} 61.5 & \cellcolor{asrcolor5} 63.5 & \cellcolor{asrcolor0} 10.0 & \cellcolor{asrcolor1} 15.0 & \cellcolor{asrcolor3} 33.5 & \cellcolor{asrcolor5} 66.0 \\
                GPTFuzzer     & \cellcolor{asrcolor7} 92.5 & \cellcolor{asrcolor7} 87.5 & \cellcolor{asrcolor1} 23.0 & \cellcolor{asrcolor6} 85.0 & \cellcolor{asrcolor1} 15.5 & \cellcolor{asrcolor7} 88.5 & \cellcolor{asrcolor0} 9.0  & \cellcolor{asrcolor0} 1.0  & \cellcolor{asrcolor5} 64.0 & \cellcolor{asrcolor4} 56.0 \\
                Avg     & \cellcolor{asrcolor4} 59.2 & \cellcolor{asrcolor4} 56.6 & \cellcolor{asrcolor3} 41.6 & \cellcolor{asrcolor5} 61.9 & \cellcolor{asrcolor2} 35.1 & \cellcolor{asrcolor5} 61.6 & \cellcolor{asrcolor1} 16.8  & \cellcolor{asrcolor1} 23.2  & \cellcolor{asrcolor1} 23.6 & \cellcolor{asrcolor5} 61.8 \\
                \bottomrule
            \end{tabular}

               \begin{flushleft}
                \textbf{Model Names:} 
                \textbf{GPT-3.5}: GPT-3.5-Turbo; 
                \textbf{Llama}: Meta-Llama-3-8B-Instruct; 
                \textbf{Qwen}: Qwen2.5-7B-Instruct;                 \textbf{Mistral}: Mistral-7B-Instruct-v0.2; 
                \textbf{DeepSeek}: DeepSeek-v3; 
                \textbf{Qwen2.5-Max}: Qwen2.5-Max-2025-01-25; 
                \textbf{o1-mini}: o1-mini; 
                \textbf{Claude-3-7}: Claude-3-7-Sonnet-20250219; 
                \textbf{Gemini-2.0}: Gemini-2.0-Flash; 
                \textbf{Qwen3-235b}: Qwen3-235b.

                % Furthermore, we categorized the models into two groups: those released before September 2024 (referred to as "old models") and those released after (referred to as "new models").

            \end{flushleft}
        \end{threeparttable}%
    }
\end{table*}

\subsection{Evaluation Setup}
In this subsection, we outline the experimental setup of the selection of models, attacks, defenses, and judges evaluated in $\mathtt{Security\; Cube}$. The selection is based on the principle to ensure coverage across all categories and provide a comprehensive view of the jailbreak landscape. Due to main page limits, the detailed description of models (Table~\ref{tab:llm_features}), datasets, attack methods, defense methods, and judge is provided in Appendix~\ref{appendix:evaluation_setup}.

\subsection{Model Robustness Landscape}

We begin by assessing the robustness of a range of large language models (LLMs), summarized in Table~\ref{tab:asr_models_updated}, spanning releases from 2023 through 2025 and sizes from 7B to 671B parameters. Robustness is evaluated using several representative jailbreak attack methods. Table~\ref{tab:asr_models_updated} reports the attack success rate (ASR) for each method–model pair, with warmer colors indicating higher ASRs.

Our analysis reveals a significant gap between earlier and more recent generations of LLMs. Legacy models released prior to 2025, such as GPT-3.5-Turbo, Qwen-2.5-7B-Instruct, and Mistral-7B-Instruct, show ASRs exceeding 90\% under adaptive attacks, including LLM-Adaptive~\cite{adaptive} and GPTFuzzer~\cite{gptfuzzer}, indicating substantial vulnerability. In contrast, the latest models, including Claude-3.7-Sonnet, Qwen-2.5-Max, and o1-mini, reduce ASRs under these strongest attacks to near zero, underscoring the impact of recent alignment advances. When aggregated across all attack methods, o1-mini and Claude-3.7-Sonnet deliver the lowest overall average ASR (approximately 17\%), positioning them at the current frontier of robustness.

We hypothesize that these robustness improvements arise from a combination of mutually reinforcing factors: \textit{i}) Deliberative alignment via chain-of-thought refusal mechanisms, enabling models to reason about safety constraints before generating answers and to internalize sophisticated refusal logic~\cite{guan2025deliberativealignmentreasoningenables}; \textit{ii}) Extensive external red-teaming and preparedness evaluations, where models undergo adversarial stress testing and iterative hardening~\cite{o1,anthropic}; and \textit{iii}) Defense-in-depth alignment frameworks, emphasizing safety-value learning, conflict resolution under ambiguous instructions, and reliability under uncertainty~\cite{openai-alignment}. Collectively, these components constitute a qualitative leap in the robustness landscape, clearly differentiating post-2025 models from their predecessors.

Our primary evaluation leverages the well-established HarmBench dataset~\cite{harmbench}, released in early 2024. To address the possibility that robustness gains in newer models stem from exposure to HarmBench during training, we performed two complementary validation experiments, detailed in Appendix~\ref{appendix:data_exposure}. These results indicate that improvements cannot be attributed solely to dataset exposure, reinforcing the conclusion that recent alignment and safety mechanisms drive the observed gains.

\takeawayN{Recent reasoning-aligned LLMs, such as o1-mini and Claude-3.7-Sonnet, show a qualitative leap in jailbreak robustness. Their resilience arises from deliberative refusal reasoning, large-scale red-teaming, and defense-in-depth alignment, marking a shift from reactive safeguards to integrated safety reasoning.}

\openproblemN{Yet, no model achieves full jailbreak resistance: advanced multi-turn and compositional attacks still yield $>$50 \% ASR. Achieving generalizable robustness remains open, requiring scalable deliberative safety, mitigation of alignment–capability trade-offs, and theory-grounded robustness metrics beyond benchmarks.}

\subsection{Jailbreak Attack Comparisons}
\noindent{\textbf{Attack concentration}}.
Table \ref{tab:cpai} summarizes the average ASR and CIPA values for each attack. Specifically, the average ASR of an attack is obtained by averaging its ASRs across all models reported in Table \ref{tab:asr_models_updated}, while the CIPA values are computed according to the metric defined in Section \ref{sec:cubeframe}. Overall, ReNeLLM, ActorBreaker, and LLM-Adaptive emerge as the most effective strategies, achieving average success rates of 66.60\%, 61.35\%, and 57.65\%, respectively. Meanwhile, CodeAttacker (CIPA = 0.11), ActorBreaker (0.11) exhibit the lowest CIPA values, indicating that these attacks generalize well across different models and thus possess broad applicability.

Notably, the most effective jailbreaks, those with both high average ASR and low CIPA, primarily fall into two categories: strategy-based and multi-round attacks, each offering distinct advantages. Strategy-based attacks leverage human insight to exploit model-specific vulnerabilities. Many are logic-driven, guiding the model to rationalize harmful outputs through carefully constructed reasoning chains. This design makes them robust across architectures and alignment settings. However, their reliance on human creativity limits scalability. Multi-round attacks are also effective because most alignment datasets primarily contain short, single-turn interactions, providing limited coverage of complex dialogue structures. Moreover, long-context exchanges may introduce distributional shifts~\cite{actorattack}, where dialogue patterns deviate from alignment data, reducing the reliability of safety mechanisms and enabling unsafe responses in later turns.

\begin{table}[t]
    \centering
\caption{Average ASR and CIPA scores for each attack method. For ASR, warmer red shades indicate higher values ($>$55\%), reflecting stronger attack performance. For CIPA, cooler green shades indicate lower values ($<$0.12), corresponding to greater attack concentration.}
\label{tab:cpai}

\resizebox{0.75\linewidth}{!}{%
\begin{tabular}{llcc}
\toprule
\textbf{Attack Category} & \textbf{Attack Method} & \textbf{Average ASR $\uparrow$} & \textbf{CIPA ($\lambda$) $\downarrow$} \\
\midrule

Strategy & ReNeLLM        & \redcell 66.60 & \greencell 0.12 \\
Multi-round & ActorBreaker    & \redcell 61.35 & \greencell 0.11 \\
Logprob & LLM-Adaptive   & \redcell 57.65 & \redcell 0.16 \\ 
Strategy & CodeAttack     & \yellowcell 54.57 & \greencell 0.11 \\
Template & GPTFuzzer      & \yellowcell 52.24 & \yellowcell 0.14 \\
LLM & PAIR           & \yellowcell 50.19 & \greencell 0.12 \\
Strategy & PAP           & \yellowcell 48.58 & \greencell 0.12 \\
Shuffle & Flip           & \yellowcell 41.52 & \yellowcell 0.13 \\
LLM & AutoDAN-Turbo  & \yellowcell 40.17 & \greencell 0.12 \\
Flaw & Multijail      & \greencell 29.74 & \yellowcell 0.14 \\
Flaw & CipherChat    & \greencell 27.42 & \redcell 0.16 \\
Strategy& DrAttack       & \greencell 25.63 & \yellowcell 0.15 \\
Shuffle & BON            & \greencell 18.13 & \redcell 0.16 \\
\bottomrule
\end{tabular}
}
\end{table}

\noindent{\textbf{Cross-model attack transferability.}}
To evaluate cross-model transferability, we construct harmful prompts on a source model for each attack and then evaluate them on a target model. The source models include DeepSeek-v3, Claude-3.7, Qwen2.5-7B, and LLaMA-3-8B, while the target models are Gemma-3-27B-it, Qwen3-8B, and LLaMA-3.2-3B. As shown in Table~\ref{tab:transfer}, the LLM-Adaptive attack achieves the highest average transfer success rate ($\approx$43\%), indicating that exploiting consistent language-modeling behaviors, such as continuation from specific initial tokens (e.g., ``sure''), can effectively induce jailbreaks across architectures. PAP and ReNeLLM also demonstrate strong transferability, suggesting that they encode model-agnostic prompting strategies. Both employ task-diverting mechanisms that guide the model to perform an ostensibly benign task (e.g., code generation) while covertly producing harmful content. Since such ``unaligned'' regions in the latent space are shared among models, these attacks generalize effectively beyond their source architectures.

We further observe that several attacks (e.g., PAIR, Flip, and DrAttack) exhibit an average $\gamma > 1$, meaning their success rate on the target model exceeds that on the source model. This phenomenon likely arises when the source model used to craft harmful prompts is more capable or safety-aware than the target. Prompts generated under stronger alignment constraints may remain unsuccessful on the source model but can easily bypass weaker safety mechanisms in less aligned targets. These findings highlight an asymmetry in LLM jailbreak defenses: once advanced models become accessible to adversaries, weaker models without comparable alignment safeguards face elevated transfer risks.

\begin{table}[t]
    \centering
    \caption{Average transfer ASR (\%) and transfer ratio ($\gamma$) for each attack. For average transfer ASR, warmer red shades indicate higher values ($>$30\%), reflecting stronger attack transferability. For average $\gamma$, cooler green shades indicate lower values ($<$0.45), corresponding to weaker transferability.}
    \resizebox{0.7\linewidth}{!}{%
    \begin{tabular}{lcc}
        \toprule
        \textbf{Attack Method} & \textbf{Average Transfer ASR} & \textbf{Average} $\gamma$ \\
        \midrule
        LLM-Adaptive & \redcell{43.42} & \yellowcell{0.46} \\
        PAP & \redcell{35.83} & \yellowcell{0.90} \\
        ReNeLLM & \redcell{32.25} & \greencell{0.43} \\
        ActorBreaker & \redcell{30.59} & \yellowcell{0.62} \\
        AutoDAN-Turbo & \yellowcell{30.00} & \yellowcell{0.65} \\
        PAIR & \yellowcell{28.37} & \redcell{1.34} \\
        Flip & \yellowcell{22.92} & \redcell{1.32} \\
        Multijail & \yellowcell{19.15} & \redcell{0.95} \\
        CodeAttacker & \greencell{18.04} & \greencell{0.42} \\
        DrAttack & \greencell{11.42} & \redcell{2.62} \\
        GPTFuzzer & \greencell{7.84} & \greencell{0.31} \\
        BON & \greencell{4.08} & \greencell{0.42} \\
        CipherChat & \greencell{3.73} & \greencell{0.14} \\
        \bottomrule
    \end{tabular}
    }
    \label{tab:transfer}
\end{table}

% Color scale: Cooler shades (blue) correspond to lower attack success rates (ASR) and thus greater system safety, whereas warmer shades (red) signify higher ASR and consequently reduced safety.
\begin{table}[t]
    \centering
    \caption{Attack overhead for different methods on Meta-Llama-3-8B-Instruct, measured in token cost and time cost. Cooler green shades denote lower token and time costs (more computationally efficient attacks); warmer shades denote higher costs.}
    \resizebox{0.7\linewidth}{!}{%
    \begin{tabular}{lcc}
    \toprule
    & \textbf{Token Cost $\downarrow$} & \textbf{Time Cost(s) $\downarrow$} \\
    \midrule
    CodeAttacker     & \greencell{888.6}   & \greencell{14.73}  \\
    Multijail      & \greencell{1827.15} & \yellowcell{105.55} \\
    Flip           & \greencell{2405.5}  & \greencell{9.73}   \\
    CipherChat    & \greencell{3890.4}  & \greencell{29.72}  \\
    ReNeLLM        & \greencell{5682.1}  & \greencell{48.13}  \\
    DrAttack       & \yellowcell{13733.9} & \yellowcell{283.31} \\
    PAP            & \yellowcell{13789.5} & \greencell{47.81}  \\
    GPTFuzzer      & \yellowcell{15959.2} & \yellowcell{121.46} \\
    BON            & \yellowcell{28780.6} & \redcell{127.43} \\
    AutoDAN-Turbo  & \redcell{57205.0}  & \redcell{427.54} \\
    PAIR           & \redcell{74161.6}  & \yellowcell{112.48} \\
    ActorBreaker    & \redcell{81789.1}  & \redcell{335.71} \\
    LLM-Adaptive   & \redcell{444005.0} & \redcell{667.58} \\
    \bottomrule
    \label{tab:attack_cost}
    \end{tabular}
    }
\end{table}

\begin{table*}[t]
    \centering
   \caption{Attack stability scores ($\beta$) for each attack method. Here, $C$ denotes jailbreak categories, and $Q_1$ and $Q_2$ represent different questions within each category. Categories: $C_1$ chemical/biological; $C_2$ illegal; $C_3$ misinformation/disinformation; $C_4$ harmful; $C_5$ harassment/bullying; $C_6$ cybercrime/intrusion. Higher $\beta$ values indicate lower attack stability. Color scale: red cells ($\beta > 0.45$), yellow cells ($0.2 < \beta \leq 0.45$), and green cells ($\beta \leq 0.2$).}

     \resizebox{0.82\linewidth}{!}{
\begin{tabular}{lccccccccccccc}
\toprule
\textbf{Attack Method} & $(C_1,Q_1)$ & $(C_1,Q_2)$ & $(C_2,Q_1)$ & $(C_2,Q_2)$ & $(C_3,Q_1)$ & $(C_3,Q_2)$ & $(C_4,Q_1)$ & $(C_4,Q_2)$ & $(C_5,Q_1)$ & $(C_5,Q_2)$ & $(C_6,Q_1)$ & $(C_6,Q_2)$ & \textbf{Average} \\
\midrule
LLM-Adaptive & \greencell{0.00} & \greencell{0.00} & \greencell{0.00} & \greencell{0.00} & \greencell{0.00} & \yellowcell{0.40} & \greencell{0.00} & \greencell{0.00} & \greencell{0.00} & \greencell{0.00} & \greencell{0.00} & \greencell{0.00} & \greencell{0.03} \\
ActorBreaker & \greencell{0.00} & \greencell{0.00} & \greencell{0.00} & \greencell{0.00} & \redcell{0.49} & \greencell{0.00} & \greencell{0.00} & \redcell{0.49} & \yellowcell{0.30} & \redcell{0.50} & \greencell{0.00} & \greencell{0.00} & \greencell{0.15} \\
BON & \greencell{0.00} & \redcell{0.50} & \greencell{0.00} & \yellowcell{0.42} & \greencell{0.00} & \yellowcell{0.31} & \greencell{0.00} & \greencell{0.00} & \yellowcell{0.42} & \greencell{0.00} & \greencell{0.00} & \yellowcell{0.31} & \greencell{0.16} \\
Flip & \yellowcell{0.40} & \greencell{0.00} & \greencell{0.00} & \redcell{0.49} & \greencell{0.00} & \greencell{0.00} & \greencell{0.00} & \greencell{0.00} & \yellowcell{0.40} & \greencell{0.00} & \greencell{0.00} & \greencell{0.00} & \greencell{0.11} \\
PAIR & \yellowcell{0.40} & \greencell{0.00} & \redcell{0.50} & \yellowcell{0.40} & \yellowcell{0.30} & \greencell{0.00} & \yellowcell{0.40} & \yellowcell{0.30} & \redcell{0.46} & \redcell{0.49} & \redcell{0.46} & \greencell{0.00} & \yellowcell{0.31} \\
AutoDAN-Turbo & \greencell{0.00} & \redcell{0.47} & \redcell{0.47} & \greencell{0.00} & \greencell{0.00} & \greencell{0.00} & \greencell{0.00} & \greencell{0.00} & \greencell{0.00} & \greencell{0.00} & \redcell{0.50} & \greencell{0.00} & \greencell{0.12} \\
ReNeLLM & \redcell{0.46} & \yellowcell{0.40} & \redcell{0.49} & \redcell{0.49} & \redcell{0.46} & \yellowcell{0.30} & \redcell{0.50} & \redcell{0.49} & \redcell{0.46} & \yellowcell{0.40} & \redcell{0.49} & \yellowcell{0.30} & \yellowcell{0.44} \\
DrAttack & \yellowcell{0.40} & \redcell{0.50} & \greencell{0.00} & \greencell{0.00} & \greencell{0.00} & \greencell{0.00} & \greencell{0.00} & \greencell{0.00} & \greencell{0.00} & \redcell{0.49} & \greencell{0.00} & \greencell{0.00} & \greencell{0.12} \\
Multijail & \redcell{0.49} & \greencell{0.00} & \greencell{0.00} & \yellowcell{0.30} & \greencell{0.00} & \greencell{0.00} & \redcell{0.50} & \redcell{0.46} & \yellowcell{0.40} & \greencell{0.00} & \greencell{0.00} & \greencell{0.00} & \greencell{0.18} \\
CipherChat & \redcell{0.46} & \greencell{0.00} & \redcell{0.49} & \redcell{0.50} & \yellowcell{0.30} & \greencell{0.00} & \greencell{0.00} & \redcell{0.49} & \yellowcell{0.30} & \yellowcell{0.40} & \yellowcell{0.30} & \greencell{0.00} & \yellowcell{0.27} \\
CodeAttacker & \greencell{0.00} & \yellowcell{0.40} & \greencell{0.00} & \greencell{0.00} & \greencell{0.00} & \greencell{0.00} & \greencell{0.00} & \yellowcell{0.30} & \greencell{0.00} & \greencell{0.00} & \greencell{0.00} & \greencell{0.00} & \greencell{0.06} \\
PAP & \redcell{0.46} & \yellowcell{0.30} & \greencell{0.00} & \greencell{0.00} & \greencell{0.00} & \greencell{0.00} & \redcell{0.49} & \greencell{0.00} & \greencell{0.00} & \yellowcell{0.40} & \yellowcell{0.40} & \yellowcell{0.30} & \greencell{0.20} \\
GPTFuzzer & \yellowcell{0.30} & \yellowcell{0.40} & \greencell{0.00} & \redcell{0.49} & \redcell{0.46} & \yellowcell{0.30} & \greencell{0.00} & \greencell{0.00} & \redcell{0.50} & \redcell{0.46} & \greencell{0.00} & \greencell{0.00} & \yellowcell{0.24} \\

\bottomrule
\end{tabular}
}
    \label{tab:beta}
\end{table*}

\noindent{\textbf{Attack overhead.}}
Table~\ref{tab:attack_cost} reports the average token and time costs of each attack across multiple attack goals on Meta-Llama-3-8B-Instruct. Overall, token costs vary substantially due to differences in attack mechanisms. LLM-Adaptive incurs the highest cost because its gradient-based, white-box optimization process requires extensive token sampling, often hundreds of thousands of tokens, to refine prompts. In contrast, Flip and CodeAttacker, which are static and non-generative (involving only template-based formatting), incur negligible low costs, making them suitable for stealthy or low-resource attack scenarios. A mid-cost group, including AutoDAN-Turbo and PAIR, exhibits variable expenses depending on interaction depth. Notably, among these, AutoDAN-Turbo achieves higher efficiency, likely due to its strategy library or other prompt optimization mechanisms, resulting in reduced token consumption.

\noindent{\textbf{Attack stability.}}
To assess attack stability, we select six common jailbreak goals, each represented by two questions. For every question, ten harmful prompts are generated using the evaluated attack method. As shown in Table~\ref{tab:beta}, LLM-Adaptive exhibits the highest stability, with an average $\beta$ of 0.03, attributed to its white-box, deterministic prompt generation process. This method consistently produces outputs initiated by fixed tokens (e.g., ``Sure''), which minimizes contextual variance. CodeAttacker also demonstrates strong stability due to its template-based design and limited prompt randomness. In contrast, ReNeLLM ($\beta = 0.44$) and PAIR ($\beta = 0.31$), although seeming to be effective with high averaged ASR in Table~\ref{tab:cpai}, show the lowest stability, with performance highly sensitive to contextual changes. This instability arises because both rely on LLM-generated prompts with weak structural constraints and unpredictable search paths, PAIR modifies prompts randomly without clear direction (e.g., identifying the model's weaknesses), while ReNeLLM relies on the model to handle diverse tasks, but frequent prompt changes hinder its ability to perform consistently. Other attack methods maintain moderate stability, performing more consistently when confined to specific harmful goals.

\noindent{\textbf{Attack depth.}} To understand how attacks affect model internals, we compare the cosine similarity of hidden states between successful and failed prompts for each attack method. As shown in Figure~\ref{fig:cossim} (Appendix~\ref{appendix:depth}), most attacks exhibit their strongest impact in the final layers, where model decision-making occurs. For example, BON, PAIR, and CipherChat induce sharp shifts near the output, suggesting direct disruption of high-level semantic representations. Others, such as multijail and GPTFuzzer, show a gradual deviation from mid to deep layers, reflecting progressive reasoning manipulation. Figures~\ref{fig:tsne5} and \ref{fig:tsne30} further illustrate how adversarial prompts alter the model's internal representations. In Figure~\ref{fig:tsne5} (early layer), successful prompts from different attack methods form distinct clusters, indicating that each attack induces a consistent activation pattern, suggesting the feasibility of automated detection via hidden states. In Figure~\ref{fig:tsne30} (deep layer), benign and jailbreak prompts are clearly separated: benign inputs cluster tightly, while all jailbreak samples shift to a distinct region. This sharp divergence highlights fundamentally different representations, which could inform future defense designs.

\takeawayN{\textbf{\textit{i}) Attack effectiveness and generality} increasingly converge: strategy-based and multi-round approaches such as ReNeLLM, ActorBreaker, and LLM-Adaptive achieve both high average ASR and low CIPA, revealing that human-crafted reasoning strategies and dialogue-structure exploitation remain the most transferable jailbreak paradigms; \textbf{\textit{ii}) Attack transferability} further exposes shared latent failure modes, linguistic continuation patterns and benign-task diversions, that persist across architectures, reflecting common generative priors rather than model-specific flaws; \textbf{\textit{iii}) Attack resource cost and stability} vary widely: white-box adaptive attacks are computationally intensive but deterministic, whereas lightweight template-based methods offer low-cost, high-stability avenues for stealthy or automated exploitation.}

\openproblemN{\textbf{\textit{i}) Unified optimization under conflicting attack objectives}. No existing jailbreak method simultaneously achieves high effectiveness, transferability, efficiency, and stability. Gains in one dimension often degrade another, exposing a lack of unified theory for multi-objective attack design. Understanding these trade-offs is essential to define the limits of optimal jailbreak capability; \textbf{\textit{ii}) Inherent vulnerability from shared generative priors}. LLMs remain intrinsically susceptible to jailbreaks due to their cooperation-oriented generative priors and autoregressive continuation bias. These shared linguistic and reasoning tendencies form systemic vulnerabilities that transcend architectures. Overcoming them demands rethinking safety at the level of generative modeling itself, not just post-hoc alignment.}

\subsection{Defense Method Comparisons}

\noindent{\textbf{Defense success rate}}.
Table~\ref{tab:defense_summary} summarizes the defense success rate of each method. Hidden State Guard is the most effective, reducing attack success to nearly zero for nine out of eleven attacks. Its high performance arises from early-stage interception of malicious prompts by analyzing the model’s internal hidden representations, enabling accurate detection and blocking before generation. System Prompt SelfReminders also substantially lower Llama’s average ASR by reinforcing safety instructions through the system prompt, though their effect on Mistral is weaker, likely because Mistral’s inherent safety alignment is less robust, limiting the prompt’s influence. Aligner (a post-filter approach) offers partial protection but are less effective against sophisticated or multi-turn attacks. CircuitBreaker (a fine-tuning-based approach) may not reduce the attack success rate (ASR) to zero but significantly lowers it for most attack methods, providing overall protection. Overall, pre-filter-based defenses, particularly those leveraging internal model states, provide the most reliable protection. In contrast, system-prompt methods depend heavily on the model’s intrinsic alignment, and post-rewriting approaches tend to be less consistent in mitigating sophisticated jailbreaks.

\begin{table}[t]
\centering
\caption{Defense success rates (\%) across different defenses. Color scale: red for high success ($>$70\%), yellow for moderate (30–70\%), and green for low ($<$30\%). Higher values indicate stronger defense performance.}
\resizebox{0.95\linewidth}{!}{
\begin{tabular}{llcccccccc}
\toprule
\multirow{2}{*}{\textbf{Attack}} & \multirow{2}{*}{\textbf{Model}} & \multicolumn{5}{c}{\textbf{Defense Method}}                                                                                                                              \\ \cmidrule{3-7}
                                 &                                 & \textbf{SelfReminder}               & \textbf{LlamaGuard}            & \textbf{Hidden State Guard}     & \textbf{Aligner}                & \textbf{CircuitBreaker}         \\ 
                                 \midrule
% \toprule
%  % &  & \multicolumn{5}{c}{\textbf{Defense method}} &  &  & \\
% \textbf{Attack} & \textbf{Model} & \textbf{Reminder} & \textbf{PromptGuard} & \textbf{Hidden State Guard} & \textbf{Aligner} & \textbf{CircuitBreaker} \\
% \midrule
\multirow{2}{*}{LLM-Adaptive} & Mistral & \greencell{16.13} & \redcell{99.44} & \redcell{100.00} & \redcell{97.76} & \redcell{89.93} \\
                             & Llama  & \redcell{96.39} & \redcell{100.0} & \redcell{100.0} & \redcell{98.97} & \redcell{89.69} \\
\cmidrule{2-7}
\multirow{2}{*}{ActorBreaker} & Mistral & \greencell{15.23} & \greencell{21.19} & \yellowcell{31.79} & \yellowcell{64.90} & \yellowcell{62.91} \\
                            & Llama  & \yellowcell{62.16} & \yellowcell{51.57} & \yellowcell{45.51} & \greencell{1.62} & \greencell{24.32} \\ 
                            \cmidrule{2-7}

\multirow{2}{*}{BON} & Mistral & \greencell{1.33} & \redcell{97.83} & \redcell{100.0} & \yellowcell{47.25} & \redcell{94.20} \\
                     & Llama  & \redcell{100.0} & \redcell{100.0} & \redcell{100.0} & \redcell{72.73} & \redcell{90.91} \\ 
                     \cmidrule{2-7}

\multirow{2}{*}{Flip} & Mistral & \greencell{0.0} & \redcell{96.09} & \redcell{100.0} & \yellowcell{39.35} & \redcell{70.65} \\
                      & Llama  & \redcell{89.74} & \redcell{94.87} & \redcell{100.0} & \greencell{5.13} & \redcell{76.92} \\ 
                      \cmidrule{2-7}

\multirow{2}{*}{PAIR} & Mistral & \greencell{15.09} & \greencell{24.53} & \redcell{96.23} & \greencell{11.95} & \yellowcell{46.54} \\
                      & Llama  & \redcell{78.38} & \greencell{0.0} & \redcell{97.3} & \greencell{14.86} & \redcell{76.44} \\ 
                      \cmidrule{2-7}

\multirow{2}{*}{AutoDAN-Turbo} & Mistral & \yellowcell{52.98} & \yellowcell{31.48} & \redcell{99.43} & \greencell{24.50} & \redcell{82.85} \\
                              & Llama  & \redcell{98.75} & \greencell{0.0} & \redcell{100.0} & \redcell{61.25} & \redcell{90.0} \\ 
                              \cmidrule{2-7}

\multirow{2}{*}{ReNeLLM} & Mistral & \yellowcell{37.36} & \yellowcell{59.34} & \redcell{99.45} & \yellowcell{31.86} & \yellowcell{68.68} \\
                         & Llama  & \redcell{78.79} & \greencell{27.27} & \redcell{100.0} & \redcell{58.33} & \yellowcell{60.61} \\ 
                         \cmidrule{2-7}

\multirow{2}{*}{DrAttack} & Mistral & \greencell{0.0} & \yellowcell{56.07} & \redcell{100.0} & \yellowcell{32.88} & \greencell{23.02} \\
                          & Llama  & \redcell{100.0} & \redcell{67.42} & \redcell{100.0} & \greencell{8.77} & \redcell{96.76} \\ 
                          \cmidrule{2-7}

\multirow{2}{*}{Multijail} & Mistral & \yellowcell{77.98} & \redcell{87.42} & \greencell{20.32} & \yellowcell{33.62} & \redcell{86.27} \\
                           & Llama  & \redcell{94.68} & \yellowcell{72.34} & \yellowcell{48.94} & \yellowcell{51.15} & \yellowcell{78.07} \\ \cmidrule{2-7}

\multirow{2}{*}{CipherChat} & Mistral & \yellowcell{55.07} & \redcell{89.71} & \redcell{100.0} & \yellowcell{57.14} & \yellowcell{59.56} \\
                            & Llama  & \redcell{100.0} & \greencell{21.95} & \redcell{100.0} & \greencell{0.0} & \redcell{100.0} \\ 
                            \cmidrule{2-7}

\multirow{2}{*}{CodeAttacker} & Mistral & \yellowcell{60.96} & \redcell{95.83} & \redcell{100.0} & \yellowcell{58.22} & \redcell{80.14} \\
                              & Llama  & \redcell{93.04} & \redcell{93.04} & \redcell{100.0} & \yellowcell{31.30} & \greencell{28.7} \\ 
                              \cmidrule{2-7}

\multirow{2}{*}{PAP} & Mistral & \greencell{1.25} & \yellowcell{33.40} & \redcell{100.0} & \greencell{1.67} & \yellowcell{67.56} \\
                     & Llama  & \redcell{93.59} & \yellowcell{35.90} & \redcell{100.0} & \yellowcell{33.33} & \redcell{96.15} \\ 
                     \cmidrule{2-7}
               
\multirow{2}{*}{GPTFuzzer} & Mistral & \redcell{70.27} & \redcell{98.25} & \redcell{98.83} & \redcell{84.73} & \redcell{97.03} \\
                         & Llama  & \redcell{93.48} & \redcell{95.65} & \redcell{100.0} & \redcell{71.74} & \redcell{86.96} \\
\bottomrule
\end{tabular}
}
\label{tab:defense_summary}
\end{table}

\begin{table}[t]
\centering
\caption{Defense overhead of different defenses.}
\resizebox{1\linewidth}{!}{
\begin{tabular}{lccccc}

\toprule
          \textbf{Overhead}            & \textbf{CircuitBreaker} & \textbf{SelfReminder} & \textbf{Hidden State Guard} & \textbf{LlamaGuard} & \textbf{Aligner} \\ \midrule
Token      &        0        &     87.01     &              298.33       &    635.69         &   2278.53      \\
Memory (MB) & 0              & 0        & 5904.22            & 15316.51    & 5904.23 \\
Latency (s)   &   0.21           &     1.36     &           3.23         &       0.82      &  29.68       \\ \bottomrule
\end{tabular}
}
 \label{tab:defend_overhead}
\end{table}

\begin{figure*}[t]
  \centering
  \begin{subfigure}[t]{0.28\textwidth}
    \centering
    \includegraphics[width=\linewidth]{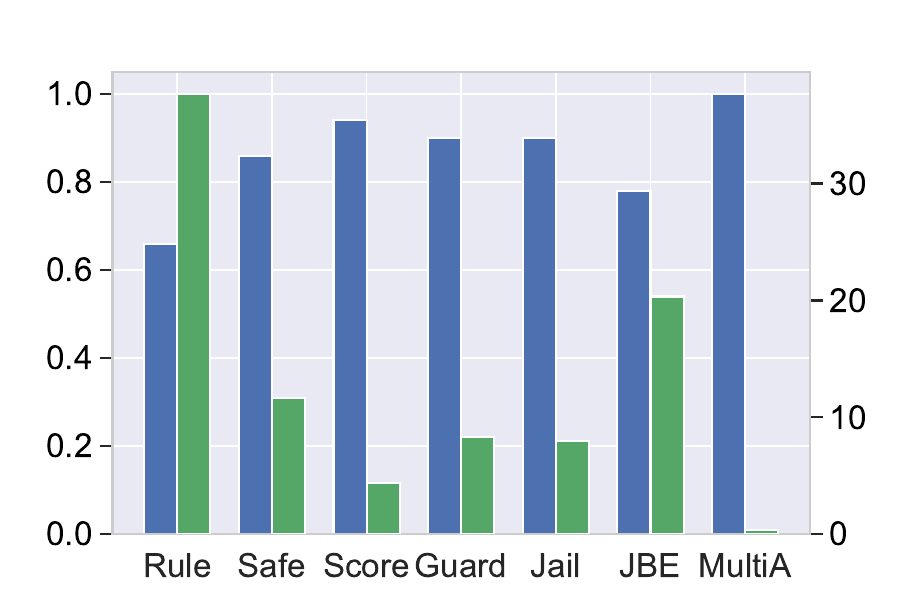}
    \caption{Accuracy (F1) and $\Delta$ASR.}
    \label{fig:Judge_F1}
  \end{subfigure}
  \hspace{0.02\textwidth}
  \begin{subfigure}[t]{0.28\textwidth}
    \centering
    \includegraphics[width=\linewidth]{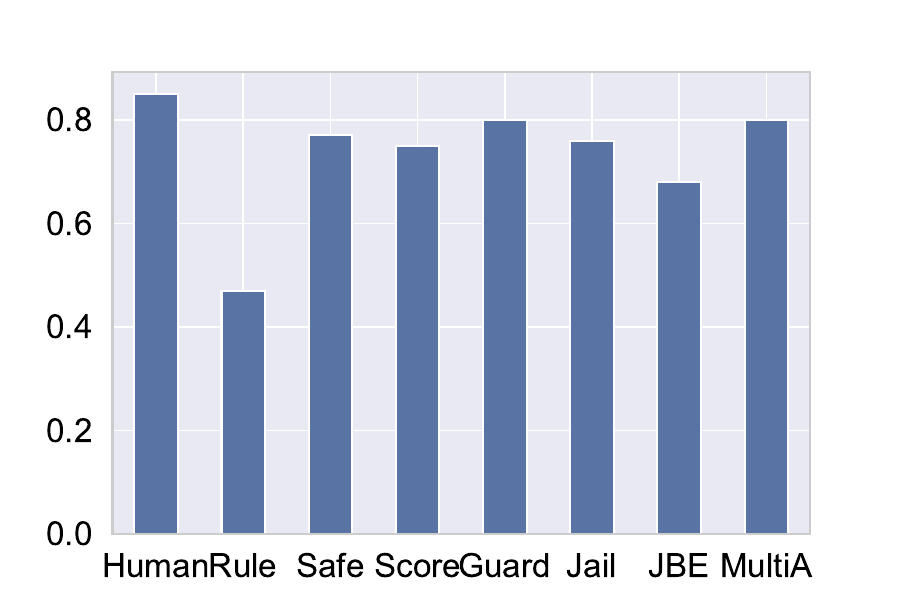}
    \caption{Fleiss's Kappa Agreement.}
    \label{fig:Judge_KP}
  \end{subfigure}
  \hspace{0.02\textwidth}
  \begin{subfigure}[t]{0.28\textwidth}
    \centering
    \includegraphics[width=\linewidth]{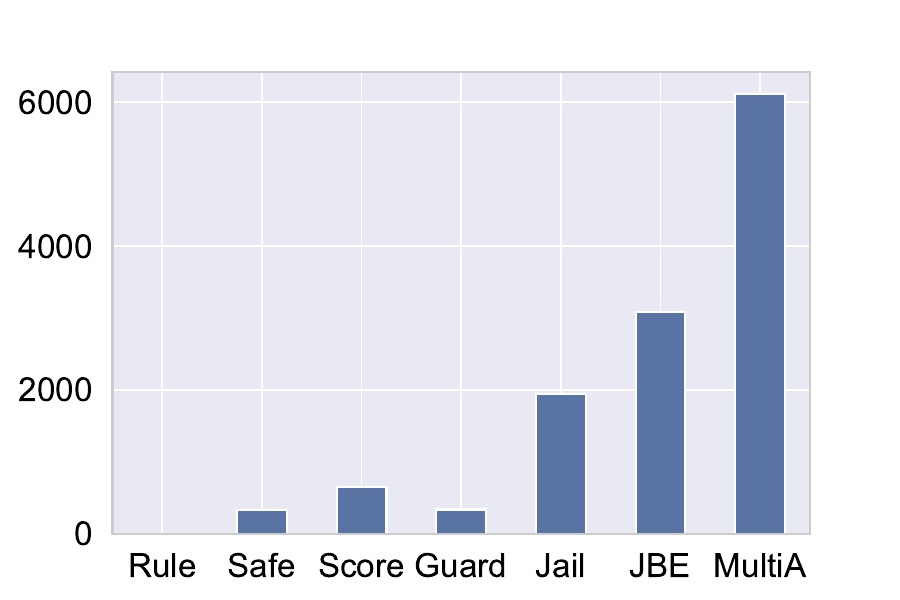}
    \caption{Evaluation Cost.}
    \label{fig:Judge_Cost}
  \end{subfigure}
  \caption{Comparison of Judge methods: Rule-based judge (Rule), binary Prompt-based (Binary Judge) (Safe), Prompt-based (Score Judge) (Score), LlamaGuard (Guard), JailJudge (Jail), JailbreakEvaluation (JBE), and Multi-Agent judge (MultiA). (a) Accuracy and alignment with human labels: Blue bars represent F1 score (higher is better) and green bars represent delta ASR (lower is better). (b) Agreement among automated judges and humans: Higher values indicate better agreement. (c) Cost in tokens: Lower values are better.}
  \label{fig:judge_all}
\end{figure*}

\begin{figure}[t]
    \centering
    \includegraphics[width=0.6\linewidth]{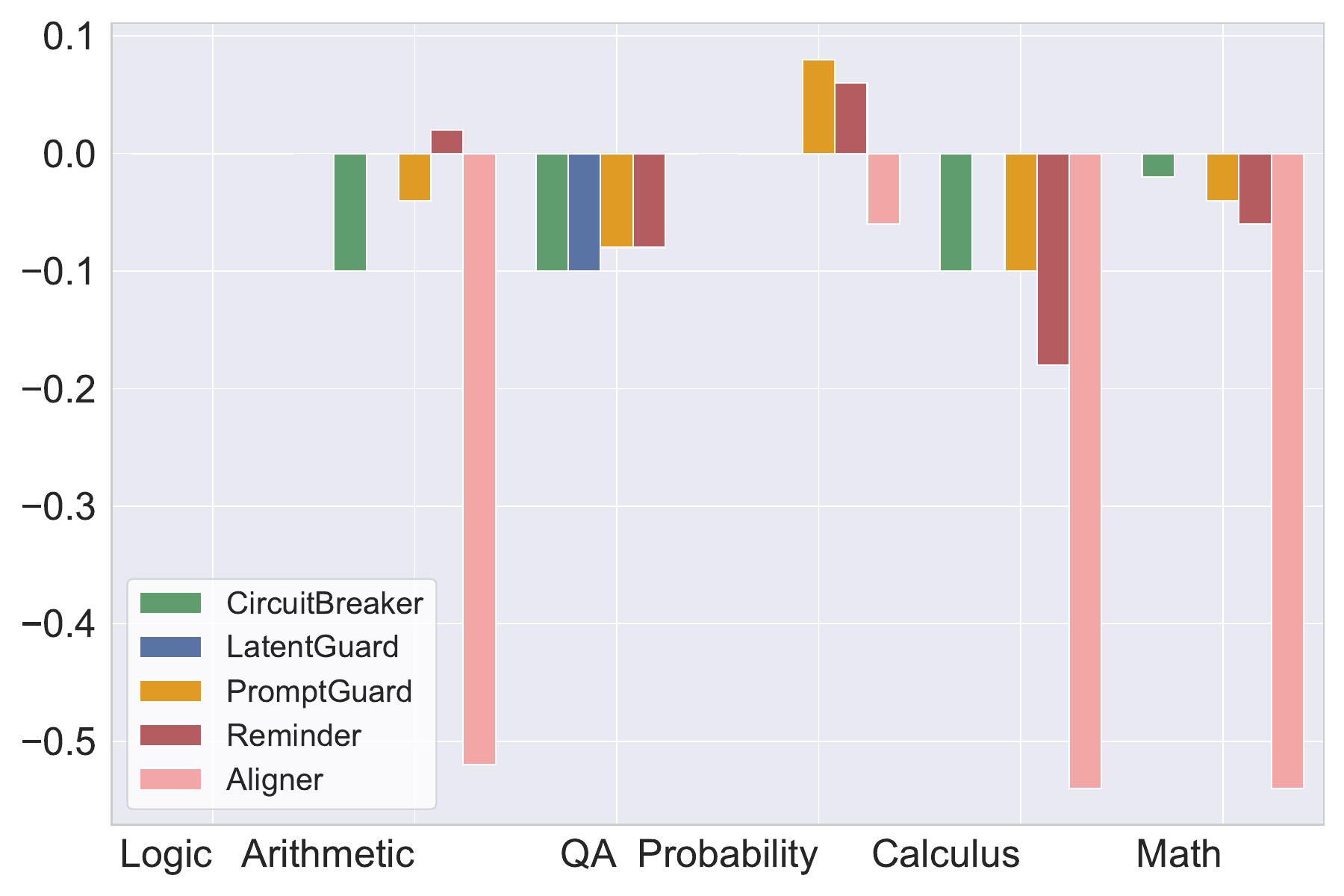}
    \caption{Utility change of each defense, measured as the percentage-point difference in task performance relative to the base model (Meta-Llama-3-8B-Instruct). Downward bars indicate reduced model utility after applying the defense. X-axis tasks: Logic, propositional logic; Arithmetic, arithmetic problems; QA, NFL question answering; Probability, probability tasks; Calculus, calculus problems; Math, graduate-level math tasks.
    }
    \label{fig:utility}
\end{figure}

\noindent \textbf{Defense overhead.}
Table~\ref{tab:defend_overhead} compares the token, memory, and latency overhead across defenses. The fine-tuned CircuitBreaker incurs virtually no additional cost, as it leverages pre-existing model behavior without extra processing. SelfReminder introduces negligible overhead ($<$50 tokens per round) through lightweight prompt edits, with minimal latency as the input length increases slightly but memory remains unchanged. Hidden State Guard and LlamaGuard add a single extra model pass for safety checks, yielding minor resource increases; the former is more memory-efficient due to its smaller 7B model but incurs extra latency from an additional classifier pass, whereas the latter performs the check directly. In contrast, Aligner is the most resource-intensive, often exceeding 4K tokens due to its multi-stage rewriting process, which requires multiple output generations. Despite using a 7B model, its runtime remains slower, likely reflecting inefficiencies in model-structure optimization.

\noindent{\textbf{Defense-utility trade-off}}.
As shown in Figure~\ref{fig:utility}, most defenses have minimal impact on math and logic tasks. Hidden State Guard, SelfReminder, and LlamaGuard cause little to no accuracy drop on benign prompts, as the first two only filter prompts that pass, meaning the utility remains unaffected, and the system prompt does not harm the model's performance if the question is benign. The fine-tuned CircuitBreaker exhibits a small accuracy drop due to capability degradation from fine-tuning, but this has been largely mitigated, as the fine-tuning process preserves the model's original ability across different tasks through careful loss function design. In contrast, Aligner induces performance declines due to answer rewriting, which can distort correct responses.

\takeawayN{\textbf{\textit{i}) Pre-filter works best}: Pre-filters based guards (e.g., Hidden State Guard and LlamaGuard) stop attacks early with minimal cost; \textbf{\textit{ii}) System prompts are lightweight but limited}: SelfReminder adds little overhead but depends on strong base alignment; best used alongside fine-tuned safety; \textbf{\textit{iii}) Post-generation fixes are inefficient}: Methods like Aligner are slow and distort outputs, while early detection is both safer and cheaper.}

\openproblemN{Combining complementary defenses, such as fine-tuning for stronger base alignment, followed by runtime filters like Hidden State Guard or SelfReminder to catch residual threats, could yield broader and more reliable protection. Evaluating such hybrid pipelines presents a promising avenue for future research.}

\subsection{Judge Method Comparisons}\label{sec:EvalJudgeCompare}

% \noindent\textbf{Agreement.}
Figure~\ref{fig:Judge_F1} shows that Multi-Agent judge align best with human judgments, achieving the lowest $\Delta$ASR (0.34\%) and highest F1 scores(0.99). Their multi-perspective design captures both surface-level and nuanced violations. Prompt-based (Score Judge) also outperforms prompt-based (Binary Judge: 11.66\%). Fine-tuned models like LlamaGuard (8.33\%) and JailJudge (8.00\%) offer moderate alignment. Figure~\ref{fig:Judge_KP} shows that LlamaGuard and Multi-Agent yield the highest inter-annotator agreement ($\kappa$ = 0.80), close to human level (0.84), indicating greater stability and reproducibility. 

In cases where LLM judges diverge from human judgment, errors often arise from over-reliance on superficial lexical cues (e.g., ``sure,'' ``here is a tutorial'') or misclassification of harmful content embedded in fictional contexts (e.g., movie). These weaknesses parallel behaviors exploited in red-teaming. In contrast, Multi-Agent and ensemble approaches mitigate such issues by integrating diverse evaluative perspectives, enabling cross-validation and more consistent judgments.

Figure~\ref{fig:Judge_Cost} shows that LlamaGuard offers the best accuracy–cost trade-off, running 47$\times$ faster and 18$\times$ cheaper (in tokens) than Multi-Agent while maintaining strong performance (F1 = 0.90, $\kappa$ = 0.80). The Multi-Agent Judge incurs higher cost due to its multi-round evaluation involving reasoning, guidance, instruction, answer, and goal assessments. In contrast, fine-tuned models like LlamaGuard require only input, answer, and goal evaluation, generating far fewer tokens. Prompt-based judges lie between these extremes, using limited guidance input but no iterative reasoning.

\takeawayN{Multi-Agent achieves the highest alignment with human labels, albeit at a higher computational cost.}

\openproblemN{Future judge designs could adopt a multi-expert architecture, where specialized evaluators assess content from distinct dimensions, such as harm potential, privacy risk, and ethical soundness, and a layered arbitration mechanism integrates these perspectives to enhance robustness, interpretability, and trustworthiness. However, it's crucial to explore the most effective sampling of perspectives and whether repeated cross-checking between agents is necessary, as without this, the cost of using multiple agents could become too high.}
\section{Outlook}

\noindent \textbf{Advancing LLM robustness through attack-defense co-evolution.}
Our analysis in experiments reveals that achieving jailbreak robust in LLMs requires moving beyond single-turn defenses and shallow alignment. First, future research should prioritize multi-turn and compositional robustness, as advanced attacks (e.g., ActorBreaker) exploit dialogue context and can bypass static guardrails. Second, representation-level defenses, such as monitoring hidden states or using circuit-breaker mechanisms, show promise for early detection and intervention, but need to be evaluated against adaptive, representation-aware attacks. Third, the research community would benefit from automated, evolving red-teaming pipelines and richer benchmarks (e.g., JailbreakBench, HarmBench) that track not only ASR but also stability, transferability, and resource cost. Fourth, reasoning-aligned safety, where models are trained to deliberate about safety, not just follow rules, emerges as a key direction, especially as models become more agentic and tool-using. Finally, as jailbreak research outpaces defenses, the community should invest in living benchmarks, coordinated disclosure, and defense-in-depth strategies that combine fine-tuning, runtime filters, and post-deployment monitoring. Together, these directions aim to close the gap between attack and defense, and advance LLM safety from ad hoc mitigation toward principled, generalizable robustness.

\noindent \textbf{Jailbreak interpretation.}
The observed effectiveness of hidden-state–based defenses suggests that adversarial prompts induce systematic and localized activation shifts in the underlying representation space, particularly within layers responsible for alignment and semantic abstraction. Investigating these activation dynamics could enable a deeper mechanistic understanding of how jailbreaks exploit model reasoning pathways. Future work should therefore aim to characterize and formalize the ``trigger'' patterns and representational signatures associated with alignment failures, integrating causal interpretability methods to trace how these internal perturbations propagate to behavioral outputs. Such an approach would bridge the gap between empirical defense design and theoretical insight, ultimately contributing to the development of models whose safety is grounded in transparent rather than post hoc output control.

\noindent \textbf{Mode internals for advanced defense.} A promising future direction lies in developing trajectory-aware defenses that reason about how prompts shape a model’s internal representations over the forward pass. Our analysis in Figure~\ref{fig:cossim} and \ref{fig:tsne} shows that jailbreaks consistently induce structured deviations in hidden-state trajectories, either abrupt shifts in the final layers or gradual drifts through mid to deep layers, revealing a geometric footprint of adversarial manipulation. Future defenses could therefore move beyond output filtering to monitor and interpret these representational trajectories, distinguishing benign from adversarial dynamics in real time. Such approaches would bridge interpretability and safety, enabling models that not only detect anomalous internal behavior but can also adaptively stabilize or correct it before harmful outputs emerge. More broadly, this line of work points toward a paradigm where robustness is enforced through representational governance, aligning model internals with safe reasoning processes rather than solely constraining surface behavior.

\noindent \textbf{Evolving safety ecosystems.} The development of adaptive, continuously evolving safety ecosystems for LLMs is essential. Rather than treating red-teaming and evaluation as static, one-off efforts, future research should pay more attention to explore automated, self-improving pipelines in which models are continually challenged by evolving adversarial strategies, e.g., ``attacker AI agent'', and retrained to strengthen alignment over time. This vision draws on ideas from adversarial self-play and automated robustness testing, offering a path toward systems that can dynamically learn from their own failures. Complementing this, the field would benefit from community-maintained benchmarks and shared knowledge infrastructures that evolve alongside emerging threats. Establishing an open, regularly updated repository of jailbreak attacks, defenses, and evaluation protocols, i.e., developed collaboratively across research groups and organizations, would provide a living benchmark for progress, preventing stagnation and fostering reproducibility.

\section{Conclusion}
This paper systematizes jailbreak attacks and defenses on large language models (LLMs) through novel taxonomies grounded in their dominant underlying mechanisms. These taxonomies reveal structural commonalities across methods and enable a comprehensive synthesis of the evolving jailbreak landscape. Building on this foundation, we critically review existing benchmarks and evaluation frameworks, identifying their limitations in scope and metric coverage. To address these gaps, we introduce the $\mathtt{Security\;Cube}$, a unified, multi-dimensional evaluation paradigm that extends beyond attack success rate to jointly characterize attacks, defenses, and judges. Using this framework, we evaluate 13 representative attacks, 5 defenses, and 4 judges under 7 attack, 3 defense, and 4 judge metrics, distilling key takeaways and open challenges.

Our analysis highlights fundamental tensions between capability, safety, and robustness, and points toward promising research directions in reasoning-aligned safety, interpretable jailbreak modeling, adaptive and efficient defenses, and a system-level safety ecosystem. Together, these directions aim to advance LLM security from ad hoc mitigation toward principled, generalizable robustness.

\section*{Acknowledgments}
We thank all the reviewers for their insightful comments. This work was partially supported by the National Natural Science Foundation of China under grants No. 62532016, No. 6257071391, and No. 62502309, the Shanghai Municipal Education Commission under grant No. ZXDF030140, and the Natural Science Foundation of Shanghai under grant No. 25ZR1402211. Shuo Wang is the corresponding author.

\section*{Ethical Considerations}
This work aims to strengthen LLM safety by systematically exposing vulnerabilities, not to promote misuse. All attacks were conducted in controlled environments, and no real-world harm was inflicted. We advocate for responsible red teaming and the open sharing of defense strategies to build safer AI systems.

\bibliographystyle{plain} % Or any other bibliography style (e.g., ieeetr, unsrt)
\bibliography{ref}  % The name of your .bib file (without the .bib extension)

\appendices

\section{Experimental Setting}\label{appendix:evaluation_setup} 

In this section, we outline the experimental setup of the selection of models, attacks, defenses, and judges evaluated in $\mathtt{Security\; Cube}$. The selection is based on the principle to ensure coverage across all categories and provide a comprehensive view of the jailbreak landscape.

\noindent\textbf{Models.} In Table~\ref{tab:llm_features}, we present a comprehensive summary of the evaluated models, spanning both open and closed systems across different sizes, release years, access modalities, and alignment approaches. This selection captures the diversity of today’s LLM landscape, ensuring broad and representative coverage for our jailbreak evaluation. The temperature is set to 0.1 for all models.

\noindent \textbf{Dataset.}
The adversarial goals in our experiments are drawn from the well-established HarmBench~\cite{harmbench} benchmark, which contains 200 distinct harmful objectives. Across all evaluated models, we executed more than 48,000 attack attempts. In each attempt, a single attack method was applied to craft a harmful prompt, which was then presented to the target model either directly or the model is applied with a defense mechanism. This large-scale evaluation produced a rich dataset of model responses, forming the basis for our subsequent analysis of jailbreak effectiveness and defense performance. Additionally, we use JailbreakBench~\cite{jbb} benchmark containing 300 questions annotated by 3 human experts to evlate the performance of different judge methods.

\noindent \textbf{Attack methods}.
We compiled a suite of 13 representative attack methods from  Table~\ref{tab:redteam_sok}, choosing one or two representatives from each category in our taxonomy to ensure broad coverage of the jailbreak space. The selected attacks are LLM-Adaptive~\cite{adaptive}, ActorBreaker~\cite{actorattack}, BON (``Best-of-N'')~\cite{bon}, Flip~\cite{flip}, PAIR~\cite{pair}, AutoDAN-Turbo~\cite{autodanturbo}, ReNeLLM~\cite{renellm}, DrAttack~\cite{drattack}, PAP~\cite{pap}, CodeAttacker~\cite{codeattack}, Multijail~\cite{multilingual}, CipherChat~\cite{cipherchat}, and GPTFuzzer~\cite{gptfuzzer}. For a fair comparison, we cap multi-turn attacks at 20 turns per trial and follow each method’s original parameter settings as reported in the respective papers.

\noindent \textbf{Defense methods}.
We select five representative defense methods (Table~\ref{tab:defense_summary}) across different categories in the taxonomy, enabling us to evaluate how well each mitigates jailbreak attacks in practice and what trade-offs they introduce. Specifically, our suite includes Hidden State Guard~\cite{hidden}, LlamaGuard~\cite{llamaguard}, SelfReminder~\cite{reminder}, Aligner~\cite{aligner}, and CircuitBreaker~\cite{circuitbreaker}. During evaluation, each defense is applied individually to isolate its standalone effectiveness, ensuring that the results clearly reflect the strengths and limitations of each approach.

\noindent \textbf{Judging methods.}
We evaluate the performance of sevenhao different judge modules. The first is the Rule-Based Judge~\cite{gcg}, which relies on predefined rules for evaluation (e.g., starting with ``I am sorry''). The second and third are both Prompt-Based LLM Judges: the first is the Score Judge, which assesses harmfulness on a scale from 1 to 10~\cite{adaptive}, and the second is the Binary (Safe) Judge~\cite{jbb}, which provides a simple success or failure judgment. The fourth and fifth are Fine-Tuned Judges, including LlamaGuard~\cite{llamaguard} and JailJudge~\cite{jailjudge}, both of which fine-tune an LLM to judge whether something is safe or not. Lastly, the Multi-Agent Judge~\cite{jailjudge} uses three agents to vote, and the Multi-Level Judge~\cite{jailbreakeval} determines jailbreak at different levels.

\begin{table}[t]
    \centering
     \caption{
    ASR of different model categories: older models (released before HarmBench) and newer models (released after HarmBench). ASR is reported on the constructed new dataset with minimal overlap with HarmBench.
    Cooler shades (green) denote low ASR (0–37.5\%, higher safety), yellow indicates moderate ASR (37.5–62.5\%), and warmer shades (red) denote high ASR (62.5–100\%, lower safety).
}
\label{tab:new_dataset}
    \resizebox{\linewidth}{!}{%
        \begin{threeparttable}
            \begin{tabular}{l *{4}{S[table-format=2.1]} | *{5}{S[table-format=2.1]}}
    \toprule
    & \multicolumn{4}{c}{\textbf{Older Models}} & \multicolumn{5}{c}{\textbf{Newer Models}} \\
                \cmidrule(lr){2-5} \cmidrule(lr){6-10}
                \textbf{Attack Method} &
                \multicolumn{1}{c}{\textbf{GPT-3.5}} & \multicolumn{1}{c}{\textbf{Llama}} & \multicolumn{1}{c}{\textbf{Qwen}} & \multicolumn{1}{c}{\textbf{Mistral}} &
                \multicolumn{1}{c}{\textbf{DeepSeek}} & \multicolumn{1}{c}{\textbf{Qwen2.5-Max}} & \multicolumn{1}{c}{\textbf{o1-mini}} & \multicolumn{1}{c}{\textbf{Claude-3-7}} & \multicolumn{1}{c}{\textbf{Qwen3-235b}} \\
                \midrule
                llmadaptive  & \cellcolor{asrcolor5} 74.0 & \cellcolor{asrcolor7} 96.0 & \cellcolor{asrcolor7} 96.0 & \cellcolor{asrcolor7} 94.0 & \cellcolor{asrcolor7} 98.0 & \cellcolor{asrcolor5} 74.0 & \cellcolor{asrcolor0} 0.0 & \cellcolor{asrcolor0} 8.0 & \cellcolor{asrcolor7} 94.0 \\
                actorattack  & \cellcolor{asrcolor5} 68.0 & \cellcolor{asrcolor6} 78.0 & \cellcolor{asrcolor6} 84.0 & \cellcolor{asrcolor6} 82.0 & \cellcolor{asrcolor6} 80.0 & \cellcolor{asrcolor6} 80.0 & \cellcolor{asrcolor6} 76.0 & \cellcolor{asrcolor4} 50.0 & \cellcolor{asrcolor6} 82.0 \\
                bon          & \cellcolor{asrcolor2} 32.0 & \cellcolor{asrcolor0} 6.0 & \cellcolor{asrcolor0} 6.0 & \cellcolor{asrcolor3} 48.0 & \cellcolor{asrcolor2} 34.0 & \cellcolor{asrcolor0} 12.0 & \cellcolor{asrcolor0} 10.0 & \cellcolor{asrcolor0} 10.0 & \cellcolor{asrcolor0} 10.0 \\
                flip         & \cellcolor{asrcolor2} 34.0 & \cellcolor{asrcolor1} 18.0 & \cellcolor{asrcolor4} 58.0 & \cellcolor{asrcolor1} 22.0 & \cellcolor{asrcolor6} 82.0 & \cellcolor{asrcolor3} 38.0 & \cellcolor{asrcolor0} 2.0 & \cellcolor{asrcolor1} 14.0 & \cellcolor{asrcolor5} 70.0 \\
                pair         & \cellcolor{asrcolor4} 58.0 & \cellcolor{asrcolor4} 54.0 & \cellcolor{asrcolor5} 74.0 & \cellcolor{asrcolor4} 56.0 & \cellcolor{asrcolor6} 78.0 & \cellcolor{asrcolor4} 62.0 & \cellcolor{asrcolor2} 36.0 & \cellcolor{asrcolor2} 32.0 & \cellcolor{asrcolor5} 74.0 \\
                autodanturbo & \cellcolor{asrcolor4} 60.0 & \cellcolor{asrcolor5} 66.0 & \cellcolor{asrcolor6} 86.0 & \cellcolor{asrcolor7} 88.0 & \cellcolor{asrcolor6} 86.0 & \cellcolor{asrcolor5} 76.0 & \cellcolor{asrcolor5} 66.0 & \cellcolor{asrcolor1} 18.0 & \cellcolor{asrcolor5} 72.0 \\
                renellm      & \cellcolor{asrcolor4} 56.0 & \cellcolor{asrcolor3} 44.0 & \cellcolor{asrcolor6} 78.0 & \cellcolor{asrcolor6} 86.0 & \cellcolor{asrcolor5} 72.0 & \cellcolor{asrcolor5} 66.0 & \cellcolor{asrcolor3} 48.0 & \cellcolor{asrcolor2} 36.0 & \cellcolor{asrcolor6} 84.0 \\
                drattack     & \cellcolor{asrcolor2} 32.0 & \cellcolor{asrcolor1} 14.0 & \cellcolor{asrcolor1} 20.0 & \cellcolor{asrcolor4} 54.0 & \cellcolor{asrcolor4} 50.0 & \cellcolor{asrcolor4} 58.0 & \cellcolor{asrcolor1} 20.0 & \cellcolor{asrcolor1} 24.0 & \cellcolor{asrcolor4} 54.0 \\
                multijail    & \cellcolor{asrcolor3} 44.0 & \cellcolor{asrcolor2} 30.0 & \cellcolor{asrcolor1} 20.0 & \cellcolor{asrcolor1} 24.0 & \cellcolor{asrcolor0} 10.0 & \cellcolor{asrcolor0} 6.0 & \cellcolor{asrcolor0} 6.0 & \cellcolor{asrcolor0} 2.0 & \cellcolor{asrcolor1} 14.0 \\
                cipherchat   & \cellcolor{asrcolor1} 14.0 & \cellcolor{asrcolor1} 16.0 & \cellcolor{asrcolor2} 32.0 & \cellcolor{asrcolor3} 46.0 & \cellcolor{asrcolor5} 72.0 & \cellcolor{asrcolor0} 12.0 & \cellcolor{asrcolor0} 4.0 & \cellcolor{asrcolor2} 28.0 & \cellcolor{asrcolor5} 66.0 \\
                codeattacker & \cellcolor{asrcolor1} 24.0 & \cellcolor{asrcolor2} 32.0 & \cellcolor{asrcolor2} 28.0 & \cellcolor{asrcolor3} 38.0 & \cellcolor{asrcolor3} 46.0 & \cellcolor{asrcolor2} 26.0 & \cellcolor{asrcolor1} 18.0 & \cellcolor{asrcolor1} 22.0 & \cellcolor{asrcolor3} 44.0 \\
                pap          & \cellcolor{asrcolor4} 52.0 & \cellcolor{asrcolor1} 14.0 & \cellcolor{asrcolor4} 62.0 & \cellcolor{asrcolor5} 70.0 & \cellcolor{asrcolor5} 68.0 & \cellcolor{asrcolor4} 50.0 & \cellcolor{asrcolor3} 48.0 & \cellcolor{asrcolor1} 20.0 & \cellcolor{asrcolor4} 54.0 \\
                gptfuzzer    & \cellcolor{asrcolor4} 54.0 & \cellcolor{asrcolor0} 2.0 & \cellcolor{asrcolor2} 30.0 & \cellcolor{asrcolor4} 62.0 & \cellcolor{asrcolor0} 12.0 & \cellcolor{asrcolor0} 2.0 & \cellcolor{asrcolor0} 6.0 & \cellcolor{asrcolor1} 18.0 & \cellcolor{asrcolor6} 75.0 \\
                \bottomrule
            \end{tabular}
        \end{threeparttable}%
    }
    \label{tab: A}
\end{table}

\begin{table}[t]
\centering
\caption{
ASR of three model categories: original older models, fine-tuned older models (trained on HarmBench), and newer models. 
Red cells indicate the average ASR in fine-tuned older models is more than 10\% points higher than that of newer models; 
Yellow cells indicate within $\pm 10\%$ percentage points; 
Green cells indicate equal or lower ASR. As shown, newer models generally exhibit a lower average ASR than fine-tuned older models, indicating robustness improvements that extend beyond training data leakage.
}
\label{tab:exposure_simulation}
\resizebox{1\linewidth}{!}{%
\begin{tabular}{llcccccc}
\toprule
                           \multirow{2}{*}{\textbf{Dataset}}& \multirow{2}{*}{\textbf{Attack Method}}            & \multicolumn{2}{c}{\textbf{Original Older Models}} & \multicolumn{2}{c}{\textbf{Fine-tuned Older Models}} & \multicolumn{2}{c}{\textbf{Newer Models}} \\ \cmidrule(lr){3-4} \cmidrule(lr){5-6} \cmidrule(lr){7-8}
                           &                  & \textbf{Qwen}   & \textbf{Mistral}  & \textbf{Qwen} & \textbf{Mistral} & \textbf{o1-mini}   & \textbf{Claude-3.7}  \\ \midrule
\multirow{13}{*}{HarmBench} & llmadaptive      & 96.0                  & 91.0                      & \worseperf{72.0}        & \worseperf{64.0}             & 0.0       & 0.0                 \\
                           & actorattack      & 78.5                  & 76.0                      & \worseperf{60.0}        & \worseperf{58.0}             & 53.0      & 26.5                \\
                           & bon              & 20.0                  & 46.0                      & \comparableperf{4.0}    & \comparableperf{4.0}          & 3.5       & 3.0                 \\
                           & flip             & 49.0                  & 32.0                      & \worseperf{28.0}        & \worseperf{2.0}              & 0.0       & 5.0                 \\
                           & pair             & 51.0                  & 96.0                      & \worseperf{50.0}        & \worseperf{44.0}              & 16.5      & 12.0                \\
                           & autodanturbo     & 52.0                  & 81.0                      & \worseperf{64.0}        & \worseperf{64.0}             & 19.5      & 37.0                \\
                           & renellm          & 82.0                  & 85.0                      & \worseperf{62.0}   & \worseperf{70.0}        & 68.5      & 79.5                \\
                           & drattack         & 37.0                  & 22.0                      & \comparableperf{12.0}   & \comparableperf{2.0}             & 2.5       & 2.0                 \\
                           & multijail        & 35.0                  & 48.0                      & \comparableperf{12.0}   & \comparableperf{10.0}         & 2.0       & 17.0                \\
                           & cipherchat       & 14.0                  & 47.0                      & \comparableperf{38.0}        & \comparableperf{2.0}              & 0.5       & 42.5                \\
                           & codeattack       & 56.0                  & 48.0                      & \betterperf{18.0}    & \betterperf{26.0}        & 33.0      & 40.0                \\
                           & pap              & 75.0                  & 80.0                      & \worseperf{42.0}         & \worseperf{20.0}         & 10.0      & 32.0                \\
                           & gptfuzzer        & 87.5                  & 74.5                      & \comparableperf{6.0}     & \comparableperf{0.0}              & 9.0       & 0.0                 \\ \midrule
\multirow{13}{*}{New Dataset} & llmadaptive      & 88.0                  & 94.0                      & \worseperf{64.0}        & \worseperf{92.0}             & 0.0       & 8.0                 \\
                           & actorattack      & 76.0                  & 82.0                      & \comparableperf{58.0}   & \comparableperf{70.0}        & 76.0      & 50.0                \\
                           & bon              & 52.0                  & 48.0                      & \comparableperf{4.0}         & \comparableperf{2.0}              & 10.0      & 10.0                \\
                           & flip             & 38.0                  & 22.0                      & \comparableperf{2.0}         & \comparableperf{8.0}         & 2.0       & 14.0                \\
                           & pair             & 92.0                  & 56.0                      & \worseperf{44.0}    & \worseperf{52.0}             & 36.0      & 32.0                \\
                           & autodanturbo     & 74.0                  & 88.0                      & \worseperf{64.0}        & \worseperf{84.0}             & 36.0      & 18.0                \\
                           & renellm          & 86.0                  & 86.0                      & \worseperf{70.0}        & \worseperf{84.0}             & 48.0      & 36.0                \\
                           & drattack         & 22.0                  & 2.0                       & \betterperf{2.0}        & \betterperf{0.0}             & 20.0      & 24.0                \\
                           & multijail        & 52.0                  & 24.0                      & \comparableperf{10.0}    & \comparableperf{2.0}             & 6.0       & 2.0                 \\
                           & cipherchat       & 35.0                  & 46.0                      & \comparableperf{2.0}         & \comparableperf{0.0}              & 4.0       & 28.0                \\
                           & codeattack       & 40.0                  & 38.0                      & \worseperf{26.0}        & \worseperf{46.0}             & 18.0      & 22.0                \\
                           & pap              & 76.0                  & 70.0                      & \betterperf{20.0}        & \betterperf{32.0}         & 48.0      & 20.0                \\
                           & gptfuzzer        & 78.0                  & 62.0                      & \betterperf{0.0}         & \betterperf{10.0}         & 6.0       & 18.0                \\ \bottomrule
\end{tabular}
}
\label{tab:B}
\end{table}

For our experiments, we primarily use the GPT-4o Score Judge as the final evaluation method. To ensure the robustness of our results, we also employ LlamaGuard, a fine-tuned model known for its effectiveness in detecting jailbreaks while maintaining reasonable time and token costs. In cases where there is disagreement between the judges, human experts are engaged to resolve the discrepancies and assess the consistency and reliability of the judgments. All reviewers are undergraduate students majoring in computer science, who were given the same instructions as those used in HarmBench for evaluating harmful content. Prior to their participation, they were informed of potential risks, including exposure to offensive content, and were ensured that all data would remain anonymized and used solely for research purposes.

\section{Training Data Exposure}\label{appendix:data_exposure}

We conducted two experiments to examine the training data exposure problem: 

\begin{itemize}[leftmargin=*]
    \item \textbf{Unseen dataset evaluation.} We constructed a new dataset of jailbreak goals designed to have minimal overlap with those in HarmBench (verified via lexical and semantic similarity filtering). This dataset allows us to assess model robustness on tasks that they are highly unlikely to have encountered during training.
    \item \textbf{Exposure simulation.} We fine-tuned two legacy models released before HarmBench, i.e., Qwen2.5-7B and Mistral-7B, directly on HarmBench, aiming to test whether deliberate exposure to benchmark data could reproduce the robustness observed in newer models.
\end{itemize}

Tables~\ref{tab:new_dataset} and~\ref{tab:exposure_simulation} present the results. We find that:

\begin{itemize}[leftmargin=*]
    \item On the new dataset, newer models still achieve lower ASR, though with slightly smaller gains (36.0\%) than on HarmBench (40.2\%), suggesting that robustness improvements generalize beyond the benchmark.
    \item Fine-tuned older models show reduced ASR on both HarmBench ($-27.9$\% on average) and the new dataset ($-26.1$\%), confirming that exposure contributes to robustness. However, their absolute ASR (32.1\%) remains substantially higher than that of top newer models (19.8\%) on HarmBench.
\end{itemize}

These findings indicate that while training exposure to benchmark data can enhance robustness, the substantial performance gap between fine-tuned legacy models and the latest models suggests that recent robustness gains primarily stem from improved data curation, multi-stage alignment, and red-teaming processes rather than benchmark leakage.

However, among these newer models, we observe that DeepSeek-v3 and Qwen-3-235B still achieve ASRs as high as 63\%, indicating that not all recent releases exhibit guaranteed robustness against jailbreaks. This disparity may stem from several factors, including incomplete adversarial coverage during safety fine-tuning, residual capability, safety trade-offs where highly capable models can inadvertently reason their way around alignment constraints, and insufficient defense-in-depth adaptation for multi-turn or compositional interactions. Put differently, model scale and stronger reasoning capabilities do not inherently ensure safety. Nonetheless, even the most robust models, such as o1-mini and Claude-3.7-Sonnet, remain susceptible to advanced multi-turn or structured attacks (e.g., ActorBreaker~\cite{actorattack}, ReNeLLM~\cite{renellm}), with ASRs reaching approximately 53--80\%. These findings highlight that while modern alignment has substantially improved baseline robustness, lingering architectural and behavioral vulnerabilities persist, underscoring that current alignment methods alone cannot fully eliminate jailbreak risks.

\section{Depth of Disruption}\label{appendix:depth}
\begin{figure}[t]
    \centering
    \includegraphics[width=0.5\linewidth,]{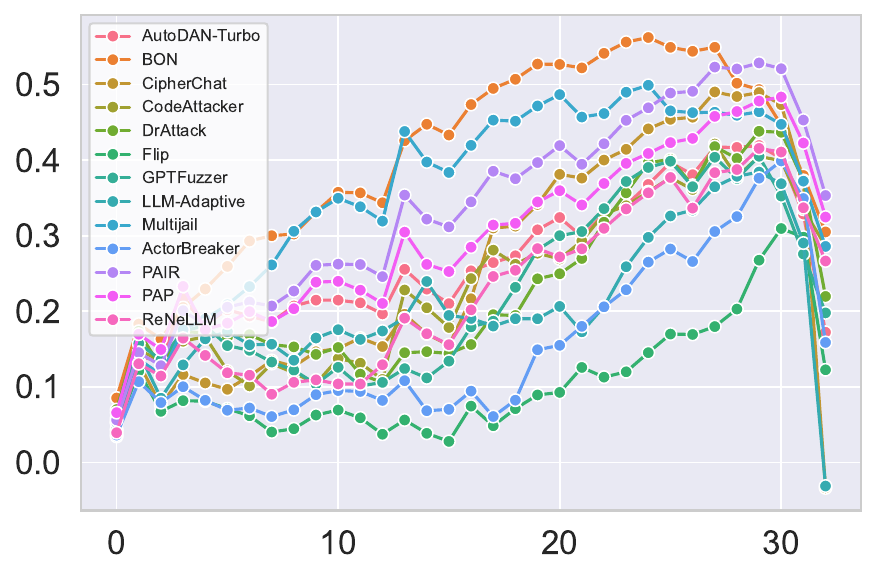}
    \caption{Cosine similarity between successful and failed jailbreak prompts in different model layers. The score fluctuates early, then drops sharply at the last layer, indicating strong separation in the model's final output space.}
    \label{fig:cossim}
\end{figure}

\begin{figure}[t]
  \centering
  \begin{subfigure}[t]{0.18\textwidth}
    \centering
    \includegraphics[width=\linewidth]{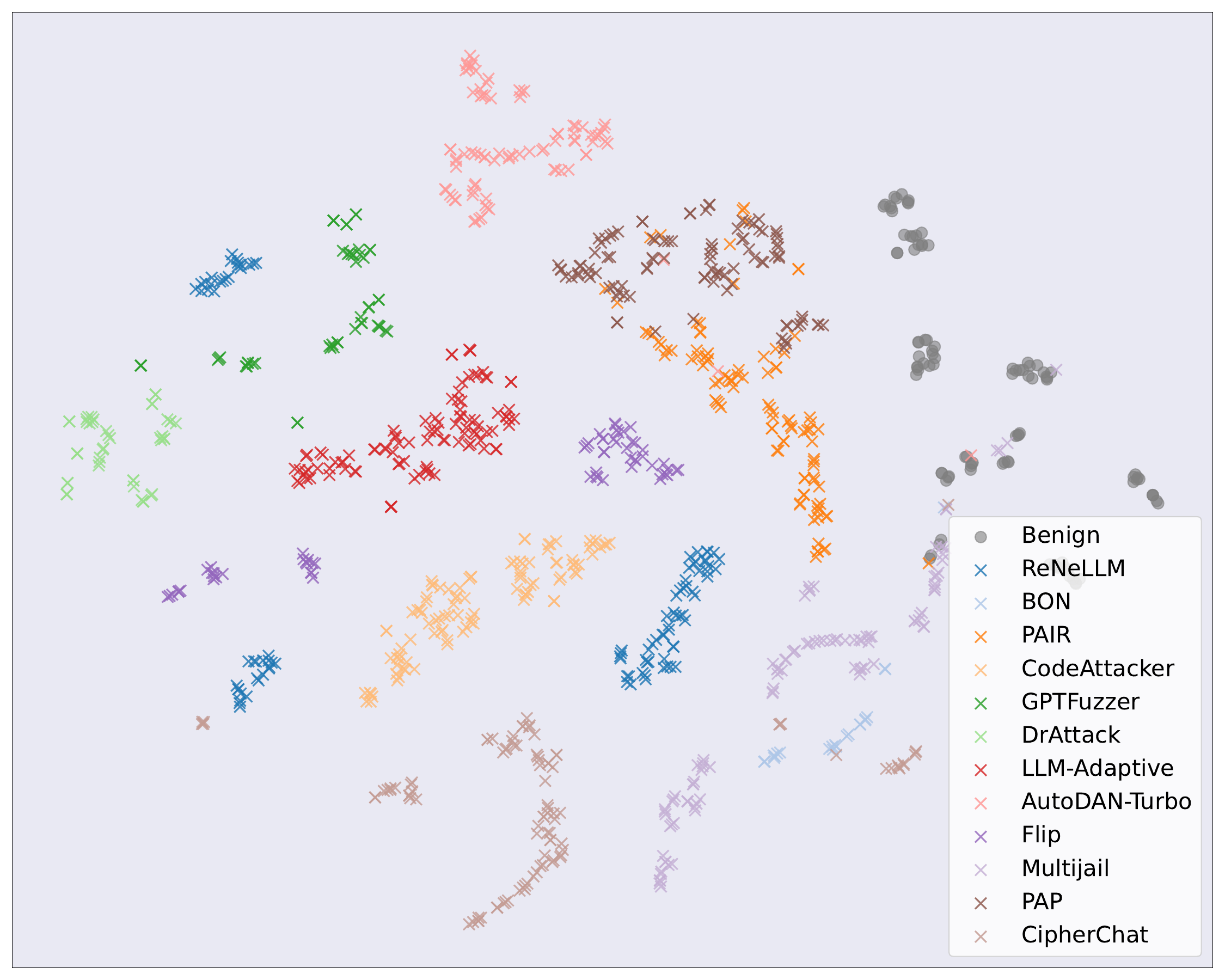}
    \caption{Layer 5.}
    \label{fig:tsne5}
  \end{subfigure}
  \hspace{0.03\textwidth}
  \begin{subfigure}[t]{0.18\textwidth}
    \centering
    \includegraphics[width=\linewidth]{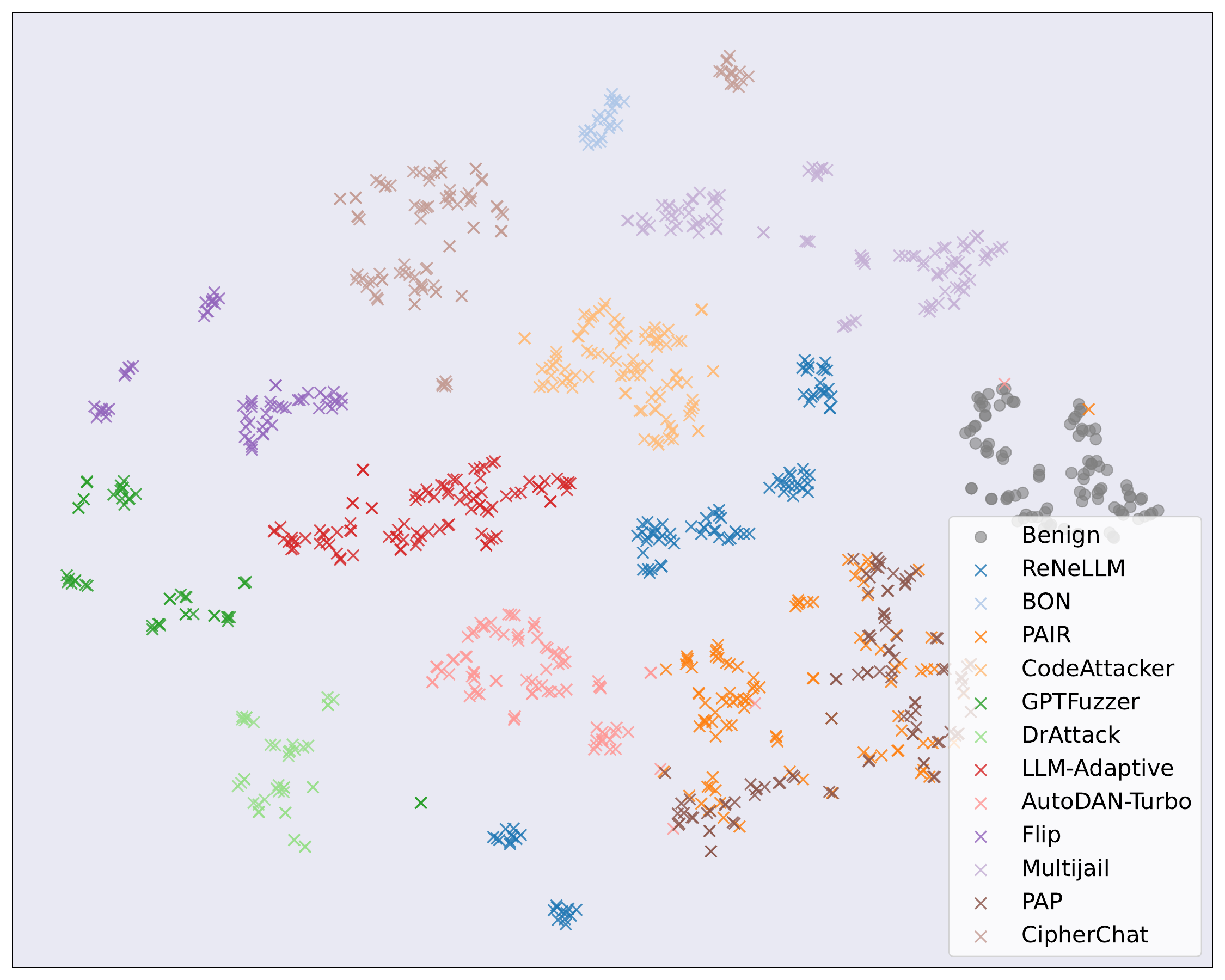}
    \caption{Layer 30.}
    \label{fig:tsne30}
  \end{subfigure}
  \caption{t-SNE visualization of successful attack prompt embeddings at Layer 5 (a) and Layer 30 (b).}
  \label{fig:tsne}
\end{figure}

To study depth of disruption in attacks, we provide the figures of cosine similarity of hidden states between successful and failed prompts for each attack method and t-SNE visualization of successful prompts in Figure~\ref{fig:cossim} and~\ref{fig:tsne}.

\newpage % The Meta-Review should at least start on a new column

% Use \appendices and not \appendix due to IEEEtran.cls quirks
%\appendices % if not used earlier

\section{Meta-Review}

The following meta-review was prepared by the program committee for the 2026
IEEE Symposium on Security and Privacy (S\&P) as part of the review process as
detailed in the call for papers.

\subsection{Summary}

This paper systematizes jailbreak attacks against LLMs. It introduces a taxonomy of attack strategies and defense mechanisms, summarizing recent research directions. Furthermore, it proposes a structured evaluation framework and presents a systematic analysis of representative attacks and defenses.

\subsection{Scientific Contributions}
\begin{itemize}
\item Provides a Valuable Step Forward in an Established Field
\end{itemize}

\subsection{Reasons for Acceptance}

\begin{enumerate}
\item The paper provides a valuable step forward in an established field by systematizing both the landscape of jailbreak attacks and corresponding defense mechanisms.
\item It introduces a structured evaluation framework (``security cube'') that advances how jailbreaks and defenses are assessed.
\item  It offers a comprehensive and systematic evaluation of representative jailbreak attacks and defenses, serving as a strong reference point for future work.
\end{enumerate}

% that's all folks
\end{document}